\documentclass[3p,times]{elsarticle}
\pdfoutput=1
\usepackage{amsmath}
\usepackage{paralist}
\usepackage{hyperref}
\hypersetup{urlcolor=blue, citecolor=red}
\usepackage{amssymb}
\usepackage{epsfig}
\usepackage{enumitem}
\usepackage{graphicx}
\usepackage{subfig}
\usepackage{float}
\usepackage{mathtools}
\usepackage{color}

\newcommand{\Rn}{\mathbb{R}}

\newcommand{\bA}{{\bf A}}

\newcommand{\bB}{{\bf B}}

\newcommand{\bF}{{\bf F}}
\newcommand{\bH}{{\bf H}}
\newcommand{\bI}{{\bf I}}

\newcommand{\bk}{{\bf k}}
\newcommand{\bK}{{\bf K}}
\newcommand{\bM}{{\bf M}}
\newcommand{\bn}{{\bf n}}
\newcommand{\bo}{{\bf 0}}

\newcommand{\bu}{{\bf u}}
\newcommand{\bU}{{\bf U}}

\newcommand{\bV}{{\bf V}}

\newcommand{\bW}{{\bf W}}
\newcommand{\bx}{{\bf x}}

\renewcommand{\Re}{\operatorname{Re}}

\newcommand{\Omt}{{{\bf \Omega}_t}}
\newcommand{\Omo}{{\bf \Omega}_0}
\newcommand{\Omht}{{{\bf \Omega}_{h,t}}}
\newcommand{\domt}{{{\bf \partial \Omega}_t}}
\newcommand{\domht}{{{\bf \partial \Omega}_{h,t}}}
\newcommand{\bbeta}{\boldsymbol{\beta}}
\newcommand{\sigmav}{\boldsymbol{\sigma}_v}
\newcommand{\sigmae}{\boldsymbol{\sigma}_e}
\newcommand{\sigmac}{\boldsymbol{\sigma}_c}
\newcommand{\sigmap}{\boldsymbol{\sigma}_p}
\newcommand{\bomega}{\boldsymbol{\omega}}
\newcommand{\beps}{\boldsymbol{\varepsilon}}
\newcommand{\bmu}{\boldsymbol{\mu}}
\newcommand{\balp}{\boldsymbol{\alpha}}

\newcommand{\bxi}{\boldsymbol{\xi}}
\newcommand{\pder}[2]{\frac{\partial#1}{\partial#2}}
\newcommand{\pderi}[1]{\frac{\partial}{\partial#1}}
\makeatletter
\def\ps@pprintTitle{%
 \let\@oddhead\@empty
 \let\@evenhead\@empty
 \def\@oddfoot{}%
 \let\@evenfoot\@oddfoot}
\makeatother
\begin{document}

\begin{frontmatter}

\title{A moving grid finite element method applied to a mechanobiochemical model for 3D cell migration.}

\author{Laura Murphy}
\ead{L.R.Murphy@sussex.ac.uk}
\author{Anotida Madzvamuse}

\address{University of Sussex, School of Mathematical and Physical Sciences,\\
	  Department of Mathematics,\\
	  BN1 9QH, UK}

\begin{abstract}
This work presents the development, analysis and numerical simulations of a biophysical model for 3D cell deformation and movement, which couples biochemical reactions and biomechanical forces. We propose a mechanobiochemical model which considers the actin filament network as a viscoelastic and contractile gel. The mechanical properties are modelled by a force balancing equation for the displacements, the pressure and concentration forces are driven by actin and myosin dynamics, and these are in turn modelled by a system of reaction-diffusion equations on a moving cell domain. The biophysical model consists of highly non-linear partial differential equations whose analytical solutions are intractable. To obtain approximate solutions to the model system, we employ the moving grid finite element method. The numerical results are supported by linear stability theoretical results close to bifurcation points during the early stages of cell migration. Numerical simulations exhibited show both simple and complex cell deformations in 3-dimensions that include cell expansion, cell protrusion and cell contraction. The computational framework presented here sets a strong foundation that allows to study more complex and experimentally driven reaction-kinetics involving actin, myosin and other molecular species that play an important role in cell movement and deformation.
\end{abstract}
\end{frontmatter}

\section{Introduction}
Cell movement is critical in multicelluar organisms due to roles in embryogenesis, wound healing, immune response, cancer metastasis, tumour invasion, and other processes, therefore, understanding cell movement is of great importance to medicine and to understanding our origins \citep{bray2001,brinkmann2004,condeelis2003,friedl2009}. In this study we consider a mechanobiochemical model previously studied by George et al. \cite{uduak2013,uduak,madzvamuse2013} which we will extend to 3-dimensions as well as introducing for the first time, the role of myosin in the modelling and computational framework. The model comprises of a system of reaction-diffusion equations for cellular proteins and a viscoelastic mechanical model for cell movement and deformation. Given that the model is highly nonlinear, exact analytical solutions are not possible to obtain in closed form, instead, we will seek to compute numerical approximations to these exact solutions. Numerical methods abound for solving complex partial differential equation (PDEs), methods that have been employed to model cell motility include finite differences, phase field methods, boundary element methods (BEM), immersed boundary methods or level set methods (LSM), \citep{alt1995,bottino1998,neilson2011,pozrikidis2005,stephanou2004,strychalski2010,wolgemuth2010}. Choosing a suitable method for a particular model is a balance between the ease of application within the model's framework and the reliability of solutions produced.

Finite differences were used in previous incarnations of the model \citep{alt1995,stephanou2004}. This method is very useful and easy to implement on fixed and simple domains but it is significantly more complicated to incorporate for the evolving domains and surfaces we wish to use and there are often problems with a moving boundary. Level set methods are used extensively in cell simulations and are useful when cells split and reconnect, therefore, it may be advantageous to use this method in the future when considering cell proliferation (cell division) and apoptosis (cell death) \citep{yang2017}. In this work we are not concerned with cells splitting. 

The finite element method is well known to easily handle complex and evolving cellular domains and can be generalised to multidimensions with little complication, hence it is the ideal method to numerically solve our model system. Finite element methods have been widely used to model cell motility \citep{blazakis2015,bottino2002,chen2012,elliot,gladilin2007,macdonald2016,manhart2017,rubinstein2005,sakamoto2014,tozluouglu2013}, and can be implemented in diverse ways depending on the model.

Given these considerations we develop a finite element based formulation which follows the work of George \cite{uduak} with the extension into multidimensions and involving solving two reaction diffusion equations. Additionally we develop our numerical solver based on deal.ii rather than ALBERTA as previously done by George \cite{uduak}.

This article is therefore structured as follows: In Section \ref{sec:model}, we introduce the mechanobiochemical model. Theoretical predictions of the spatiotemporal behaviour of the solutions of the model close to bifurcation points are presented in \ref{sec:fsgc}, these identify important bifurcation parameters. In Section \ref{sec:FEM}, theoretical predictions are used to validate finite element simulations since there are no analytical solutions to compare with. Further finite element simulations illustrating 3D cell movement and deformation are exhibited in Section \ref{sec:sim_full}. We discuss findings and conclude in Section \ref{sec:conclusion}.

\section{A mechanobiochemical model}\label{sec:model}
The model we consider and extend is inspired by contractile models of the actin cytogel \cite{lewis1991,oster1985}. These models comprise of a force balance equation modelling the displacements of the cell when deformed and a reaction-diffusion equation for the concentration of the gel that in turn drives cell movement. The idea of pressure driven protrusion and the use of concentration of actin originates from Alt and Tranquillo \cite{alt1995}. In their model they assume movement is produced by a balance between contractile force of the actin network pulling on the membrane and pressure pushing on the membrane. This was extended by Stephanou et al. so that large deformations could be modelled which is more realistic for most cells \cite{stephanou2004}. George further extended this model by observing (and hence modelling) that higher actin concentration in a region leads to more pressure \cite{uduak}. In the previous models a polar coordinate system was used and radial extension of the cell was calculated \citep{alt1995,stephanou2004}. Unlike this approach, we follow the work of George and study the mechanobiochemical model in its physical Cartesian coordinates without any need for coordinate transformation \cite{uduak}. We extend of the work by George in two key ways, firstly by extending from two, to three dimensions and secondly by adding the consideration of the concentration of myosin. We model the concentrations of, and interactions between, actin and myosin using two reaction-diffusion equations. The reaction-diffusion equations are coupled to a force balance equation which describes the movement of the cell. The model equations are outlined as follows.

We assume that the cell shape is a simply connected and continuously deforming domain: $\Omt \subset \Rn^3$ with boundary $\domt$, where $t \in I=[0,T_f],$ $ T_f>0$. Any point $\bx\in\Omt$ is defined by $\bx=(x(t),y(t),z(t))$. We define the displacement of $\bx$ at time $t$ by $\bu=(u(\bx(t),t),v(\bx(t),t),w(\bx(t),t))^T$. Let the concentration of F-actin, and bound myosin, at point $\bx(t)$ be given by $a=a(\bx(t),t)$, and $m=m(\bx(t),t)$, respectively. We describe the dynamics of the actin network by the following system of the equations,
\begin{subequations}\label{eq:fs}
\begin{align}
\nabla \cdot (\sigmav+\sigmae+\sigmac+\sigmap)&=\bo & \text{in } & \Omt, t\in I,\label{eq:cm1}\\
\pder{a}{t} + \nabla \cdot (a\bbeta) -D_a \Delta a - f(a,m) &=0  & \text{in } & \Omt, t\in I,\label{eq:cm2}\\
\pder{m}{t} + \nabla \cdot (m\bbeta) -D_m \Delta m - g(a,m) &=0  & \text{in } & \Omt, t\in I,\label{eq:cm3}\\
a(\bx(t),t) = a_0, \quad \bu(\bx(t),t) &= \bo & \text{for } & \bx\in\Omo,\\
\bbeta &= \bomega_n & \text{for } & \bx\in\domt, t\in I,\\
\sigmav \cdot \bn = \sigmae \cdot \bn = \bn \cdot \nabla a= \bn \cdot \nabla m &= 0 & \text{for } & \bx\in\domt, t\in I,\label{eq:bc}
\end{align} 
\end{subequations}
where the viscoelastic and contractile properties are described by stress tensors:
\begin{itemize}
 \item viscous $\sigmav(\bu)=\mu_1\pder{\beps}{t}+\mu_2\pder{\phi}{t}\bI$ where $\mu_1$ and $\mu_2$ are shear and bulk viscosities respectively, $\beps$ is the the strain tensor ($\frac{1}{2}(\nabla\bu+\nabla\bu^T)$) and $\phi$ is the dilation ($\nabla \cdot \bu$).
 \item elastic $\sigmae(\bu)=\frac{E}{1+v}(\beps+\frac{\nu}{1-2\nu}\phi\bI)$ where $E$ is the Youngs modulus and $\nu$ is the Poisson ratio.
 \item contractile $\sigmac(a,m)=(\psi a^2 e^{-a/a_{sat}}+cm) \bI$, where $\psi$ and $c$ are the contractility coefficients for $a$ and $m$, respectively, and $a_{sat}$ is the saturation coefficient of actin.
 \item pressure $\sigmap(\bu,a)=\dfrac{p}{1+\phi}\big(1+\frac{2}{\pi}\delta(l)\arctan a\big)\bI$. This describes two types of pressure. First is the hydrostatic pressure which is present everywhere and corresponds to the osmotic pressure in the cell which depends on the dilation $\phi$ and pressure coefficient $p$. Close to the membrane there is also polymerisation pressure caused by the polymerising actin filaments pushing on the cell membrane. This increases with increasing concentration of filaments $a$. We choose close to the membrane to mean less than 20\% of the cell radius from the edge in the initial state. To define this we use $\delta(l)$ and the points $\bxi = (\xi_x,\xi_y,\xi_z)\in\Omo$. There exists a family of bijective mappings between the initial and current domains we can let $l:\Omt\times I \to \Rn$ and corresponding $\hat{l}:\Omo\times I \to [0,1]$ then $\hat{l}(\bxi,t)=l(\bx(\bxi,t),t)$. So we calculate the distance from the centroid by
\begin{equation}
 \delta(l) = \left\{\begin{array}{l l}
1, & \; \text{if }\sqrt{\xi_x^2+\xi_y^2+\xi_z^2}>0.8,  \\
0, & \; \text{otherwise.}
\end{array} \right.
\end{equation}
\end{itemize}
In the reaction-diffusion equations we have the diffusion coefficients for actin and myosin, $D_a$ and $D_m$ respectively. Because the cell is moving we introduce the flow velocity $\bbeta=\pder{\bu}{t}$. The interactions between actin and myosin are described by the reaction terms $f(a,m)$ and $g(a,m)$ respectively. We have formulated different plausible reaction kinetics in the absence of experimental data and for the sake of brevity, we only present here one such model. We refer the interested reader to consult \cite{murphythesis2018}. Although we will discuss the implementation of one specific plausible model, other models can be easily incorporated and studied in a similar fashion. For illustrative purposes we consider the following hypothetical reaction kinetics 
\begin{subequations}\label{eq:specken}
 \begin{align}
 f(a,m) &= k_a(a_c-a)+k_{am}\frac{a^2(m_c-m)}{1+Ka^2} , \\
 g(a,m) &= -k_{ma}(a_c-a)-k_{am}\frac{a^2(m_c-m)}{1+Ka^2},
  \end{align}
\end{subequations}
where we begin with the same reaction term, $k_a(a_c-a)$, as used in \cite{uduak2013,uduak,madzvamuse2013}. $k_a$ is the rate of polymerisation/depolymerisation and $a_c$ is the equilibrium concentration and if the concentration is above this critical value then F-actin will depolymerise at the same rate. Next, since myosin binds to actin the amount of myosin will increase due to higher concentration of actin, hence the term $-k_{ma}(a_c-a)$ where $k_{ma}$ is the rate of binding/unbinding of myosin. Defining $m_c$ as the unstable equilibrium concentration of $m$, the last term in the actin equation represents that actin will depolymerise with higher concentrations of myosin and is subject to a saturation coefficient $K$, for $a$. The negation is true for myosin since myosin is seen to accumulate.

Thus we have three connected equations: the solutions to \eqref{eq:cm2} (actin concentration) and \eqref{eq:cm3} (myosin concentration) affect the contractile and pressure parts of the force balance equation and the solution to \eqref{eq:cm1} (displacement) affects the reaction-diffusion equations through the convection terms and the changing shape of the domain.

\subsection{Summary of results from linear stability theory}
The linear stability analysis detailed in \ref{sec:fsgc} reveals that parameters, in particular $\psi$ and $c$, can be varied so that particular patterns become unstable and grow. These patterns correspond to eigenfunctions of the Laplacian on the chosen volume. 
\section{Finite element formulation}\label{sec:FEM}
We have formulated a very complex non-linear system of partial differential equations. It is not possible to analytically solve this system, therefore we must turn to numerical methods to produce approximations. Previous versions of this model were represented in a polar coordinate system and solved using finite differences however George \cite{uduak2013} used a finite element formulation. We proceed similarly and describe our methods below. In particular, we employ the moving grid finite element method \citep{baines1994,madzvamuse2006,madzvamuse2003,madzvamuse2013} to compute approximate numerical solutions of the coupled viscoelastic reaction-diffusion system defined in 3D Cartesian coordinate system. 

To begin, the force balance is separated into a system of three partial differential equations representing the three space dimensions. This clarifies the derivation of the weak formulation. Since $\sigmav,\sigmae,\sigmac$ and $\sigmap$ (as described in Section \ref{sec:model}) are all stress tensors we can write them in matrix form. 
In three dimensions, strain and dilation are given by
\begin{subequations}
 \begin{align}
  \epsilon(\bu):=\frac{1}{2}(\nabla\bu+(\nabla\bu)^T)=\begin{pmatrix}\pder{u}{x} & \frac{1}{2}(\pder{v}{x}+\pder{u}{y}) & \frac{1}{2}(\pder{u}{z}+\pder{w}{x}) \\[6pt] \frac{1}{2}(\pder{v}{x}+\pder{u}{y}) & \pder{v}{y} & \frac{1}{2}(\pder{v}{z}+\pder{w}{y}) \\[6pt] \frac{1}{2}(\pder{v}{x}+\pder{u}{y}) & \frac{1}{2}(\pder{w}{y}+\pder{v}{z}) & \pder{w}{z} \end{pmatrix} \quad\text{ and }\quad \phi(\bu):=\pder{u}{x}+\pder{v}{y}+\pder{w}{z},
 \end{align}
\end{subequations}
respectively. It follows then that we can write the stress tensors in three-dimensional tensor-matrix form:
 \begin{subequations}
 \begin{align*}
&  \sigmav = \begin{pmatrix}
\mu_{1+2}\pder{\dot{u}}{x}+\mu_2(\pder{\dot{v}}{y}+\pder{\dot{w}}{z}) & \frac{\mu_1}{2}(\pder{\dot{v}}{x}+\pder{\dot{u}}{y}) & \frac{\mu_1}{2}(\pder{\dot{w}}{x}+\pder{\dot{u}}{z}) \\[10pt]
\frac{\mu_1}{2}(\pder{\dot{v}}{x}+\pder{\dot{u}}{y}) & \mu_2(\pder{\dot{u}}{x}+\pder{\dot{w}}{z})+\mu_{1+2}\pder{\dot{v}}{y} & \frac{\mu_1}{2}(\pder{\dot{v}}{z}+\pder{\dot{w}}{y}) \\[10pt]
\frac{\mu_1}{2}(\pder{\dot{w}}{x}+\pder{\dot{u}}{z}) & \frac{\mu_1}{2}(\pder{\dot{v}}{z}+\pder{\dot{w}}{y}) & \mu_{1+2}\pder{\dot{w}}{z}+\mu_2(\pder{\dot{u}}{x}+\pder{\dot{v}}{y}) 
\end{pmatrix}, \\[10pt]
&  \sigmae = \frac{E}{1+\nu}\begin{pmatrix}
\pder{u}{x}+\nu'\phi(\bu) & \frac{1}{2}(\pder{v}{x}+\pder{u}{y}) & \frac{1}{2}(\pder{u}{z}+\pder{w}{x}) \\[10pt]
\frac{1}{2}(\pder{v}{x}+\pder{u}{y}) & \pder{v}{y}+\nu'\phi(\bu) & \frac{1}{2}(\pder{v}{z}+\pder{w}{y}) \\[10pt]
\frac{1}{2}(\pder{w}{x}+\pder{u}{z}) &  \frac{1}{2}(\pder{v}{z}+\pder{w}{y}) & \pder{w}{z}+\nu'\phi(\bu) 
\end{pmatrix},\\[10pt]
&  \sigmac = \begin{pmatrix}\psi a^2 e^{-a/a_{sat}}+cm & 0 & 0 \\[6pt] 0 & \psi a^2 e^{-a/a_{sat}}+cm & 0 \\[6pt] 0 & 0 & \psi a^2 e^{-a/a_{sat}}+cm \end{pmatrix},\\[15pt]
&  \sigmap = \begin{pmatrix}\frac{p}{1+\phi}\left(1+\frac{2}{\pi}\delta(l)\tan^{-1} a\right) & 0 & 0 \\[6pt] 0 & \frac{p}{1+\phi}\left(1+\frac{2}{\pi}\delta(l)\tan^{-1} a\right) & 0 \\[6pt] 0 & 0 & \frac{p}{1+\phi}\left(1+\frac{2}{\pi}\delta(l)\tan^{-1} a\right)\end{pmatrix},
 \end{align*}
\end{subequations}
where $\mu_{1+2}=\mu_1+\mu_2$ and $\nu'=\nu/(1-2\nu)$. Substituting these expressions into $\nabla \cdot (\sigmav+\sigmae+\sigmac+\sigmap)=\bo$ gives us three equations

\begin{subequations}
 \begin{align*}
  \pderi{x}\Bigg(D_{11}\pder{\dot{u}}{x}+D_{12}\bigg(\pder{\dot{v}}{y}+\pder{\dot{w}}{z}\bigg)+C_{11}\pder{u}{x}+C_{12}\bigg(\pder{v}{y}+\pder{w}{z}\bigg)\Bigg)\hspace{5cm}&\\+\pderi{y}\Bigg(D_{33}\bigg(\pder{\dot{v}}{x}+\pder{\dot{u}}{y}\bigg)+C_{33}\bigg(\pder{v}{x}+\pder{u}{y}\bigg)\Bigg) + \pderi{z}\Bigg(D_{33}\bigg(\pder{\dot{w}}{x}+\pder{\dot{u}}{z}\bigg)+C_{33}\bigg(\pder{w}{x}+\pder{u}{z}\bigg)\Bigg)&=-\pder{f_1}{x}, \\ \\
  \pderi{x}\Bigg(D_{33}\bigg(\pder{\dot{v}}{x}+\pder{\dot{u}}{y}\bigg)+C_{33}\bigg(\pder{v}{x}+\pder{u}{y}\bigg)\Bigg)+\pderi{y}\Bigg(D_{11}\pder{\dot{v}}{y}+D_{12}\bigg(\pder{\dot{u}}{x}+\pder{\dot{w}}{z}\bigg)\hspace{4cm}&\\+C_{11}\pder{v}{y}+C_{12}\bigg(\pder{u}{x}+\pder{w}{z}\bigg)\Bigg) + \pderi{z}\Bigg(D_{33}\bigg(\pder{\dot{w}}{y}+\pder{\dot{w}}{z}\bigg)+C_{33}\bigg(\pder{w}{z}+\pder{v}{z}\bigg)\Bigg)&=-\pder{f_2}{y},\\ \\
  \pderi{x}\Bigg(D_{33}\bigg(\pder{\dot{w}}{x}+\pder{\dot{u}}{z}\bigg)+C_{33}\bigg(\pder{w}{x}+\pder{u}{z}\bigg)\Bigg)+\pderi{y}\Bigg(D_{33}\bigg(\pder{\dot{v}}{z}+\pder{\dot{w}}{y}\bigg)+C_{33}\bigg(\pder{v}{z}+\pder{w}{y}\bigg)\Bigg) \hspace{3cm}&\\+ \pderi{z}\Bigg(D_{11}\pder{\dot{w}}{z}+D_{12}\bigg(\pder{\dot{v}}{y}+\pder{\dot{u}}{x}\bigg)+C_{11}\pder{w}{z}+C_{12}\bigg(\pder{u}{x}+\pder{v}{y}\bigg)\Bigg)&=-\pder{f_3}{z},
 \end{align*}
\end{subequations}
where
\begin{subequations}
 \begin{gather}
  f_1=f_2=f_3=\left[\frac{p}{1+\phi}\left(1+\frac{2}{\pi}\delta(l)\arctan a\right)+\psi a^2 e^{-a/a_{sat}}+cm\right],\\
  D_{11}=\mu_1+\mu_2,\quad D_{12}=\mu_2, \quad D_{33}=\frac{\mu_1}{2},\\
  C_{11}=\frac{E(1-\nu)}{(1+\nu)(1-2\nu)}, \quad C_{12}=\frac{E\nu}{(1+\nu)(1-2\nu)}\text{ and } C_{33}=\frac{E}{2(1+\nu)}.
 \end{gather}
\end{subequations}

\subsection{Weak formulation}
To find the weak formulation, we take the usual route and multiply by a test function $\hat{\phi}(\bx,t)\in H^1(\Omt)$, where $H^1(\Omt)$ is a Hilbert space, and integrate over the domain. This takes into account Green's formula and the boundary conditions. The boundary condition $\sigmav \cdot \bn = \sigmae \cdot \bn = 0$ means that boundary term disappears during integration. The weak formulation is to find $u(\bx,t)$, $v(\bx,t)$ and $w(\bx,t)\in H^1(\Omt)$, $t\in I$ such that
\begin{subequations}
 \begin{align*}
    \int_\Omt &\pder{\hat{\phi}}{x}\left(D_{11}\pder{\dot{u}}{x}+D_{12}\bigg(\pder{\dot{v}}{y}+\pder{\dot{w}}{z}\bigg)+C_{11}\pder{u}{x}+C_{12}\bigg(\pder{v}{y}+\pder{w}{z}\bigg)\right)+\pder{\hat{\phi}}{y}\Bigg(D_{33}\bigg(\pder{\dot{v}}{x}+\pder{\dot{u}}{y}\bigg)\\&+C_{33}\bigg(\pder{v}{x}+\pder{u}{y}\bigg)\Bigg) + \pder{\hat{\phi}}{z}\left(D_{33}\bigg(\pder{\dot{w}}{x}+\pder{\dot{u}}{z}\bigg)+C_{33}\bigg(\pder{w}{x}+\pder{u}{z}\bigg)\right)d\Omt=-\int_\Omt\pder{\hat{\phi}}{x}f_1 d\Omt+\int_\domt\hat{\phi} f_1 n_1 ds,\\ \\
  \int_\Omt &\pder{\hat{\phi}}{x}\left(D_{33}\bigg(\pder{\dot{v}}{x}+\pder{\dot{u}}{y}\bigg)+C_{33}\bigg(\pder{v}{x}+\pder{u}{y}\bigg)\right)+\pder{\hat{\phi}}{y}\Bigg(D_{11}\pder{\dot{v}}{y}+D_{12}\bigg(\pder{\dot{u}}{x}+\pder{\dot{w}}{z}\bigg)\\ &+C_{11}\pder{v}{y}+C_{12}\bigg(\pder{u}{x}+\pder{w}{z}\bigg)\Bigg) + \pder{\hat{\phi}}{z}\left(D_{33}\bigg(\pder{\dot{w}}{y}+\pder{\dot{w}}{z}\bigg)+C_{33}\bigg(\pder{w}{z}+\pder{v}{z}\bigg)\right)d\Omt =-\int_\Omt\pder{\hat{\phi}}{y}f_2 d\Omt+\int_\domt\hat{\phi} f_2 n_2 ds,\\ \\
  \int_\Omt &\pder{\hat{\phi}}{x}\left(D_{33}\bigg(\pder{\dot{w}}{x}+\pder{\dot{u}}{z}\bigg)+C_{33}\bigg(\pder{w}{x}+\pder{u}{z}\bigg)\right)+\pder{\hat{\phi}}{y}\left(D_{33}\bigg(\pder{\dot{v}}{z}+\pder{\dot{w}}{y}\bigg)+C_{33}\bigg(\pder{v}{z}+\pder{w}{y}\bigg)\right) \\& +\pder{\hat{\phi}}{z}\left(D_{11}\pder{\dot{w}}{z}+D_{12}\bigg(\pder{\dot{v}}{y}+\pder{\dot{u}}{x}\bigg)+C_{11}\pder{w}{z}+C_{12}\bigg(\pder{u}{x}+\pder{v}{y}\bigg)\right)d\Omt=-\int_\Omt\pder{\hat{\phi}}{z}f_3 d\Omt+\int_\domt\hat{\phi} f_3 n_3 ds.
 \end{align*}
\end{subequations}
\noindent Since $\pder{f_1}{x}, \pder{f_2}{y}$ and $\pder{f_3}{z}$ are difficult to evaluate, we have used identities derived from the gradient theorem to write the weak form as above. In other words we have used the identity
\begin{equation}
\int_\Omt\pder{f_j}{x}\hat{\phi} d\Omt = -\int_\Omt\pder{\hat{\phi}}{x}f_j d\Omt+\int_\domt\hat{\phi} f_j n_j ds,
\end{equation}
for $j=1,2,3$, where $x$ can also be substituted by $y$ and $z$. $n_1,\;n_2,\;n_3$ are the direction cosines of the outward unit vector $\bn$ normal to $\domt$.

Next we want to find the weak formulation of the reaction-diffusion equations which are given as
\begin{equation}
 \pder{a}{t}+ \nabla \cdot (a\bbeta) -D_a \Delta a = f(a,m), \quad \pder{m}{t}+ \nabla \cdot (m\bbeta) -D_m \Delta m = g(a,m).
\end{equation}
We apply the product rule and convert to the material derivative (defined as $\frac{Da}{Dt}=\pder{a}{t}+a(\nabla\cdot\bbeta)$, in \cite{reddy1993}). This gives
\begin{subequations}
 \begin{align*}
   \frac{Da}{Dt}-D_a\Delta a + a(\nabla\cdot\bbeta)=f(a,m),\\ \frac{Dm}{Dt}-D_m\Delta m + m(\nabla\cdot\bbeta)=g(a,m).
 \end{align*}
\end{subequations}
Now continuing as with the force balance equation, we multiply by a test function $\hat{\psi}(\bx,t)\in H^1(\Omt)$ and integrate over the domain.
The terms $(D_a\Delta a) \hat{\psi}$ and $(D_m\Delta m) \hat{\psi}$ can be simplified using the divergence theorem and for the remaining part of the left hand side we can use the Reynolds transport theorem. This means the weak formulation can be written as: Find $a(\bx,t),\;m(\bx,t)\in H^1(\Omt), t\in I$ such that
\begin{subequations}
\begin{align}
 \pderi{t}\int_\Omt a\hat{\psi} d\Omt + \int_\Omt \Big(D_a\nabla a\cdot\nabla \hat{\psi}\Big) d\Omt=\int_\Omt \bigg(f(a,m)\hat{\psi}+a\frac{D\hat{\psi}}{Dt}\bigg)\; d\Omt,\\
 \pderi{t}\int_\Omt m\hat{\psi} d\Omt + \int_\Omt \Big(D_m\nabla m\cdot\nabla \hat{\psi}\Big) d\Omt=\int_\Omt \bigg(g(a,m)\hat{\psi}+m\frac{D\hat{\psi}}{Dt}\bigg)\; d\Omt,
 \end{align}
\end{subequations}
for all  $\hat{\psi}(\bx,t)\in H^1(\Omt)$.
\subsection{Space discretisation}
We now wish to define the problem at discrete points in space. To do this, we define the computational domain $\Omega_{h,t}$ as a polyhedral approximation to $\Omt$, $T_{h,t}$ the discretisation of $\Omega_{h,t}$ made up of non-degenerate elements $\kappa_i$ and the finite element space $V_h(t) := \{ v_h\in C^0(\Omt):v_h|_\kappa \text{ is linear}\}$. Thus the space-discrete problem is to find $u_h(\bx,t)$, $v_h(\bx,t)$, $w_h(\bx,t)$, $a_h(\bx,t)$, $m_h(\bx,t)\in V_h(t),\;t\in I$, such that

\begin{subequations}
 \begin{align*}
  \int_\Omht &\pder{\hat{\phi}}{x}\Bigg(D_{11}\pder{\dot{u}_h}{x}+D_{12}\bigg(\pder{\dot{v}_h}{y}+\pder{\dot{w}_h}{z}\bigg)+C_{11}\pder{u_h}{x}+C_{12}\bigg(\pder{v_h}{y}+\pder{w_h}{z}\bigg)\Bigg)+\pder{\hat{\phi}}{y}\Bigg(D_{33}\bigg(\pder{\dot{v}_h}{x}+\pder{\dot{u}_h}{y}\bigg)+C_{33}\bigg(\pder{v_h}{x}+\pder{u_h}{y}\bigg)\Bigg) \\
  & \hspace{3cm} +\pder{\hat{\phi}}{z}\Bigg(D_{33}\bigg(\pder{\dot{w}_h}{x}+\pder{\dot{u}_h}{z}\bigg)+C_{33}\bigg(\pder{w_h}{x}+\pder{u_h}{z}\bigg)\Bigg)d\Omt =-\int_\Omht\pder{\hat{\phi}}{x}f_1 d\Omht+\int_\domht\hat{\phi} f_1 n_1 ds,\\ \\
  \int_\Omht &\pder{\hat{\phi}}{x}\Bigg(D_{33}\bigg(\pder{\dot{v}_h}{x}+\pder{\dot{u}_h}{y}\bigg)+C_{33}\bigg(\pder{v_h}{x}+\pder{u_h}{y}\bigg)\Bigg)+\pder{\hat{\phi}}{y}\Bigg(D_{11}\pder{\dot{v}_h}{y}+D_{12}\bigg(\pder{\dot{u}_h}{x}+\pder{\dot{w}_h}{z}\bigg)+C_{11}\pder{v_h}{y}+C_{12}\bigg(\pder{u_h}{x}+\pder{w_h}{z}\bigg)\Bigg)  \\
  &\hspace{3cm} +\pder{\hat{\phi}}{z}\Bigg(D_{33}\bigg(\pder{\dot{w}_h}{y}+\pder{\dot{w}_h}{z}\bigg)+C_{33}\bigg(\pder{w_h}{z}+\pder{v_h}{z}\bigg)\Bigg)d\Omht =-\int_\Omht\pder{\hat{\phi}}{y}f_2 d\Omht+\int_\domht\hat{\phi} f_2 n_2 ds,\\ \\
  \int_\Omht &\pder{\hat{\phi}}{x}\Bigg(D_{33}\bigg(\pder{\dot{w}_h}{x}+\pder{\dot{u}_h}{z}\bigg)+C_{33}\bigg(\pder{w_h}{x}+\pder{u_h}{z}\bigg)\Bigg)+\pder{\hat{\phi}}{y}\Bigg(D_{33}\bigg(\pder{\dot{v}_h}{z}+\pder{\dot{w}_h}{y}\bigg)+C_{33}\bigg(\pder{v_h}{z}+\pder{w_h}{y}\bigg)\Bigg) \\
  &+ \pder{\hat{\phi}}{z}\Bigg(D_{11}\pder{\dot{w}_h}{z}+D_{12}\bigg(\pder{\dot{v}_h}{y}+\pder{\dot{u}_h}{x}\bigg)+C_{11}\pder{w_h}{z}+C_{12}\bigg(\pder{u_h}{x}+\pder{v_h}{y}\bigg)\Bigg)d\Omht=-\int\Omht\pder{\hat{\phi}}{z}f_3 d\Omht+\int_\domht\hat{\phi} f_3 n_3 ds,
 \end{align*}
\end{subequations}
and 
\begin{subequations}
 \begin{align*}
 \pderi{t}\int_{\Omega_{h,t}} a_h\hat{\psi} d\Omega_{h,t} + \int_{\Omega_{h,t}} D_a\nabla a_h\cdot &\nabla \hat{\psi} d{\Omega_{h,t}}=\int_{\Omega_{h,t}} \Bigg(I_hf(a_h,m_h)\hat{\psi}+a_h\frac{D\hat{\psi}}{Dt}\Bigg)\; d\Omega_{h,t},\\
 \pderi{t}\int_{\Omega_{h,t}} m_h\hat{\psi} d\Omega_{h,t} + \int_{\Omega_{h,t}} D_m\nabla m_h\cdot &\nabla \hat{\psi} d{\Omega_{h,t}}=\int_{\Omega_{h,t}} \Bigg(I_hg(a_h,m_h)\hat{\psi}+m_h\frac{D\hat{\psi}}{Dt}\Bigg)\; d\Omega_{h,t},
 \end{align*}
\end{subequations}
for all $\hat{\phi},\;\hat{\psi}\in V_h(t)$. We can then express $u_h,v_h,w_h,a_h$ and $m_h$ in terms of the linear basis functions:
\begin{equation}
 u_h=\sum_{j=1}^{nde}U_j\phi_j,\quad v_h=\sum_{j=1}^{nde}V_j\phi_j,\quad w_h=\sum_{j=1}^{nde}W_j\phi_j, \quad a_h=\sum_{j=1}^{nde}\alpha_j\phi_j\quad\text{ and }\quad m_h=\sum_{j=1}^{nde}\mu_j\phi_j.
\end{equation}
This means that we are left with equations which contain only simple functions and their derivatives and point values for the variables. Substituting $u_h$, $v_h$, and $w_h$ into the above equations
and using the Galerkin formulation, take the test functions to belong the same spaces as the nodal basis functions. Hence, the force balance equations can be written in block matrix-vector form
\begin{equation}
 \begin{bmatrix} \bA^{11} & \bA^{12} & \bA^{13} \\ [\bA^{12}]^T & \bA^{22} & \bA^{23} \\ [\bA^{13}]^T & [\bA^{23}]^T & \bA^{33} \end{bmatrix}\begin{Bmatrix}\frac{d\bU}{dt} \\ \frac{d\bV}{dt} \\ \frac{d\bW}{dt} \end{Bmatrix} + \begin{bmatrix} \bB^{11} & \bB^{12} & \bB^{13} \\ [\bB^{12}]^T & \bB^{22} & \bB^{23} \\ [\bB^{13}]^T & [\bB^{23}]^T & \bB^{33} \end{bmatrix}\begin{Bmatrix}\bU \\ \bV \\ \bW \end{Bmatrix} =\begin{Bmatrix}\bF^1 \\ \bF^2 \\ \bF^3\end{Bmatrix},
\end{equation}
where $\{\bU(t)\}=(U_1,...,U_{nde})$, $\{\bV(t)\}=(V_1,...,V_{nde})$ , $\{\bW(t)\}=(W_1,...,W_{nde})$ and:

\begin{subequations}
 \begin{gather*}
  \bA_{ij}^{11}(t):=\int_{{\bf \Omega}_{h,t}}D_{11}\pder{\phi_i}{x}\pder{\phi_j}{x}+D_{33}\left(\pder{\phi_i}{y}\pder{\phi_j}{y}+\pder{\phi_i}{z}\pder{\phi_j}{z}\right)d{\bf \Omega}_{h,t},\\
  \bA_{ij}^{22}(t):=\int_{{\bf \Omega}_{h,t}}D_{33}\left(\pder{\phi_i}{x}\pder{\phi_j}{x}+\pder{\phi_i}{z}\pder{\phi_j}{z}\right)+D_{11}\pder{\phi_i}{y}\pder{\phi_j}{y}d{\bf \Omega}_{h,t},\\
  \bA_{ij}^{33}(t):=\int_{{\bf \Omega}_{h,t}}D_{33}\left(\pder{\phi_i}{x}\pder{\phi_j}{x}+\pder{\phi_i}{y}\pder{\phi_j}{y}\right)+D_{11}\pder{\phi_i}{z}\pder{\phi_j}{z}d{\bf \Omega}_{h,t},\\
  \bB_{ij}^{11}(t):=\int_{{\bf \Omega}_{h,t}}C_{11}\pder{\phi_i}{x}\pder{\phi_j}{x}+C_{33}\left(\pder{\phi_i}{y}\pder{\phi_j}{y}+\pder{\phi_i}{z}\pder{\phi_j}{z}\right)d{\bf \Omega}_{h,t},\\
  \bB_{ij}^{22}(t):=\int_{{\bf \Omega}_{h,t}}C_{33}\left(\pder{\phi_i}{x}\pder{\phi_j}{x}+\pder{\phi_i}{z}\pder{\phi_j}{z}\right)+C_{11}\pder{\phi_i}{y}\pder{\phi_j}{y}d{\bf \Omega}_{h,t},\\
  \bB_{ij}^{33}(t):=\int_{{\bf \Omega}_{h,t}}C_{33}\left(\pder{\phi_i}{x}\pder{\phi_j}{x}+\pder{\phi_i}{y}\pder{\phi_j}{y}\right)+C_{11}\pder{\phi_i}{z}\pder{\phi_j}{z}d{\bf \Omega}_{h,t},\\
  \bA_{ij}^{12}(t):=\int_{{\bf \Omega}_{h,t}}D_{12}\pder{\phi_i}{x}\pder{\phi_j}{y}+D_{33}\pder{\phi_i}{y}\pder{\phi_j}{x}d{\bf \Omega}_{h,t},\quad
  \bA_{ij}^{13}(t):=\int_{{\bf \Omega}_{h,t}}D_{12}\pder{\phi_i}{x}\pder{\phi_j}{z}+D_{33}\pder{\phi_i}{z}\pder{\phi_j}{x}d{\bf \Omega}_{h,t},\\
  \bA_{ij}^{23}(t):=\int_{{\bf \Omega}_{h,t}}D_{12}\pder{\phi_i}{y}\pder{\phi_j}{z}+D_{33}\pder{\phi_i}{z}\pder{\phi_j}{y}d{\bf \Omega}_{h,t},\quad
  \bB_{ij}^{12}(t):=\int_{{\bf \Omega}_{h,t}}C_{12}\pder{\phi_i}{x}\pder{\phi_j}{y}+C_{33}\pder{\phi_i}{y}\pder{\phi_j}{x}d{\bf \Omega}_{h,t},\\
  \bB_{ij}^{13}(t):=\int_{{\bf \Omega}_{h,t}}C_{12}\pder{\phi_i}{x}\pder{\phi_j}{z}+C_{33}\pder{\phi_i}{z}\pder{\phi_j}{x}d{\bf \Omega}_{h,t},\quad
  \bB_{ij}^{23}(t):=\int_{{\bf \Omega}_{h,t}}C_{12}\pder{\phi_i}{y}\pder{\phi_j}{z}+C_{33}\pder{\phi_i}{z}\pder{\phi_j}{y}d{\bf \Omega}_{h,t},\\
  \bF_j^1(t):=-\int_{\Omega_{h,t}} f_1\pder{\phi_j}{x}d\Omega_{h,t}+\int_{\partial \Omega_{h,t}} n_1f_1\phi_j ds,\quad\quad
  \bF_j^2(t):=-\int_{\Omega_{h,t}} f_2\pder{\phi_j}{y}d\Omega_{h,t}+\int_{\partial \Omega_{h,t}} n_2f_2\phi_j ds,\quad\;\;\\
  \bF_j^3(t):=-\int_{\Omega_{h,t}} f_3\pder{\phi_j}{z}d\Omega_{h,t}+\int_{\partial \Omega_{h,t}} n_3f_3\phi_j ds.\quad\;\;
 \end{gather*}
\end{subequations}

It must be noted that in the above we have used the fact that $\hat{\phi} \in V_h$ and $\hat{\psi} \in V_h$ where $V_h = {\phi_1, .... , \phi_n}$. For the sake of ease of notation and computation we we define the following block matrices and vectors
\begin{subequations}
 \begin{gather}
 [\bA] := \begin{bmatrix} \bA^{11} & \bA^{12} & \bA^{13} \\ [\bA^{12}]^T & \bA^{22} & \bA^{23} \\ [\bA^{13}]^T & [\bA^{23}]^T & \bA^{33} \end{bmatrix}, \; 
 [\bB] := \begin{bmatrix} \bB^{11} & \bB^{12} & \bB^{13} \\ [\bB^{12}]^T & \bB^{22} & \bB^{23} \\ [\bB^{13}]^T & [\bB^{23}]^T & \bB^{33} \end{bmatrix}, \;
 \{\bU\} := \begin{Bmatrix}\bU \\ \bV \\ \bW \end{Bmatrix}\text{ and } \; \{\bF\} := \begin{Bmatrix}\bF^1 \\ \bF^2 \\ \bF^3\end{Bmatrix}.
 \end{gather}
\end{subequations}
Therefore the force balance equation's semi-discrete finite element formation can be written compactly as
\begin{equation}
 [\bA]\frac{\{d\bU\}}{dt}+[\bB]\{\bU\}=\{\bF\}.
\end{equation}
Now considering the reaction kinetics to be as in Eqs \eqref{eq:specken}, (in Section \ref{sec:model}), we can similarly write the reaction-diffusion equations in semi-discrete form
\begin{subequations}
 \begin{align}
   \pderi{t} (\bM\balp)+D_a\bK\balp=k_aa_c\bH-k_a\bM\balp+k_{am}\bM\frac{\balp(1-\bmu)}{1+K\balp^2},\\
   \pderi{t} (\bM\bmu)+D_m\bK\bmu=-k_{ma}a_c\bH+k_{ma}\bM\balp-k_{am}\bM\frac{\balp(1-\bmu)}{1+K\balp^2},
 \end{align}
\end{subequations}
where vector operations in $\balp(1-\bmu)/(1+K\balp^2)$ are pointwise and 
\begin{equation}
 M_{i,j}=\int_{\Omega_{h,t}} \phi_i \phi_j, \quad K_{i,j}=\int_{\Omega_{h,t}} \nabla \phi_i\cdot\nabla\phi_j \quad \text{and} \quad H_j=\int_{\Omega_{h,t}}\phi_j.
\end{equation}
To compute these integrals we use Gauss numerical quadrature \citep{press2007}. This is done as follows. First we can consider the integrals elementwise,
\begin{equation}
 M_{i,j}=\sum_{\Delta_k}\int_{\Delta_k} \phi_i \phi_j, \quad A_{i,j}=\sum_{\Delta_k}\int_{\Delta_k} \nabla \phi_i\cdot\nabla\phi_j \quad \text{and} \quad H_j=\sum_{\Delta_k}\int_{\Delta_k}\phi_j.
\end{equation}
 Then choose a numerical quadrature comprising a set of points and weights depending on the functions to be integrated. This can be written as a formula for the integral of a function $\xi$
\begin{equation}
 \int \xi(\bx)\approx \sum_q \xi(\bar\bx_q)w_q,
\end{equation}
where $\bar\bx_q$ and $w_q$ are the $q$th quadrature points and weights respectively. Therefore the integrals can be approximated by
\begin{equation}
\begin{split}
 M_{i,j}\approx\sum_{\Delta_k}\sum_q \phi_i(\bar\bx_q) \phi_j(\bar\bx_q)w_q, \quad A_{i,j}\approx\sum_{\Delta_k}\sum_q \nabla \phi_i(\bar\bx_q)\cdot\nabla\phi_j(\bar\bx_q)w_q \\ \text{and} \quad H_j\approx\sum_{\Delta_k}\sum_q\phi_j(\bar\bx_q)w_q.
\end{split}
\end{equation}
These are all computed in the deal.II implementation \citep{dealii}.

\subsection{Time discretisation}
Next we carry out the temporal discretisation of the system of ordinary differential equations arising from the finite element discretisation. To proceed, we split the interval into a finite number of sub-intervals $[t^n,t^{n+1}]$ use a uniform time step $\Delta t:=t^{n+1}-t^n$. We can then use a modified implicit-explicit (IMEX) finite differentiation formula \citep{lakkis2013,madzvamuse2006,ruuth}. Thus the fully discrete problem is now
\begin{subequations}\label{eq:fem}
 \begin{align}
  \left([\bA]^n+\Delta t[\bB]^n\right)\bU^{n+1}=&[\bA]^n\{\bU\}^n+\Delta t\{\bF\}^n,\label{eq:fbfe}\\
   \left[\bM^{n+1}+\Delta tD_a\bK^{n+1}\right]\balp^{n+1}=
\bM^n\balp^n+&\Delta t(k_a(a_c\bH^n-\bM^n\balp^n)+k_{am}\bM^n\frac{\balp^n(1-\bmu^n)}{1+K(\balp^n)^2}),\label{eq:rdfea}\\
 \big[\bM^{n+1}+\Delta tD_m\bK^{n+1}\big]\bmu^{n+1}= \bM^n\bmu^n+\Delta t&(-k_{ma}(a_c\bH^n-\bM^n\balp^n)-k_{am}\bM^n\frac{\balp^n(1-\bmu^n)}{1+K(\balp^n)^2}),
\label{eq:rdfem}
 \end{align}
\end{subequations}
where the superscripts ${ }^n$ and ${ }^{n+1}$ are the computed values on the mesh at times $t^n$ and $t^{n+1}$ respectively. Note that we have treated some parts implicitly (e.g. diffusion) and other parts fully explicit (e.g. reactions). 

Hence we have three equations all with the same form. At each time-step we assemble the matrices to obtain a system of linear algebraic equations. When solving \eqref{eq:fbfe} we see that the block matrix on the left hand side is not symmetric therefore we use the most effective solver for this which is GMRES \citep{saad1986}. The equations \eqref{eq:rdfea} and \eqref{eq:rdfem} are solved using the conjugate gradient method \citep{CG}.

\subsection{Nodal displacements}
The displacement of the nodes of the mesh is chosen to be equal to the flow velocity therefore $\bbeta:=\pder{\bU}{t}$. Since $t^{n+1}=t^n+\Delta t$ and $\bx(t^n)\in{\bf \Omega}_{t^n},\bx(t^{n+1})\in{\bf \Omega}_{t^{n+1}}$ be points in the respective domains. We can define a first order linear approximation as:
\begin{equation}
 \bbeta(\bx,t^n)=\frac{\bx(t^{n+1})-\bx(t^n)}{\Delta t}.
\end{equation}
This means we can define a new approximation to the domain ${\bf \Omega}_{t^{n+1}}$ such that
\begin{equation}
 \bx(t^{n+1})=\bx(t^n)+\Delta t \pder{\bU}{t}=\bx(t^n)+(\bU^{n+1}-\bU^n).
\end{equation}
At each step we have a new mesh with new shape functions so we must assemble new matrices $\bM^n,\bH^n, \bA^n, \bB^n, \bF^n$ to iteratively solve the discrete coupled problem as outlined in the following algorithm
\subsection{Numerical algorithm}
The fully discrete problem is solved iteratively with the following algorithm:
\begin{itemize}
 \item Initialise $\bU^0$, $\balp^0$, $\bmu^0$ and fixed parameters 
 \item WHILE $t<endtime$
 \begin{itemize}
  \item Assemble $\bM^n,\bH^n, \bA^n, \bB^n, \bF^n$
  \item Solve for $\bU^{n+1}$ using \eqref{eq:fbfe}
  \item Compute the new domain using $\bU^{n+1}$
  \item Solve for $\balp^{n+1}$ and $\bmu^{n+1}$ using \eqref{eq:rdfea} and  \eqref{eq:rdfem}
  \item $t = t+\Delta t$
 \end{itemize}
 \item END
\end{itemize}

We create a mesh using Gmsh \citep{geuzaine2009} and implement this algorithm using deal.II \citep{dealii}, a C++ software library which provides tools to solve partial differential equations which are discretised with finite element methods. Unlike the majority of other finite element software, deal.II uses hexahedral and quadrilateral elements rather than triangles.

\section{Numerical simulations}\label{sec:sim_full}
We now present simulations of our model. We want to see the organisation of the molecular species into regions which will cause the cell to move. This organisation may be caused by diffusion-driven instability, or due to the movement of the cell combined with the reaction-diffusion equations. The linear stability analysis of \ref{sec:fsgc} holds true close to critical bifurcation points, these include parameters as well as the geometrical deformation of the cell. The conditions for stability are numerous and complex, however it is still possible to choose parameters so that particular modes can be selected. When we consider longer time, and therefore far away from equilibrium, linear stability theory no longer holds but we see significant protrusions and contractions which deform the mesh into many different shapes. Parameters used are in Table \ref{table:fs_params}.

\subsection{Excitation of mode \texorpdfstring{$w_{1,1}^1$}{w11}}\label{sec:selw11}
\begin{figure}[H]
\centering
\includegraphics[width=12cm]{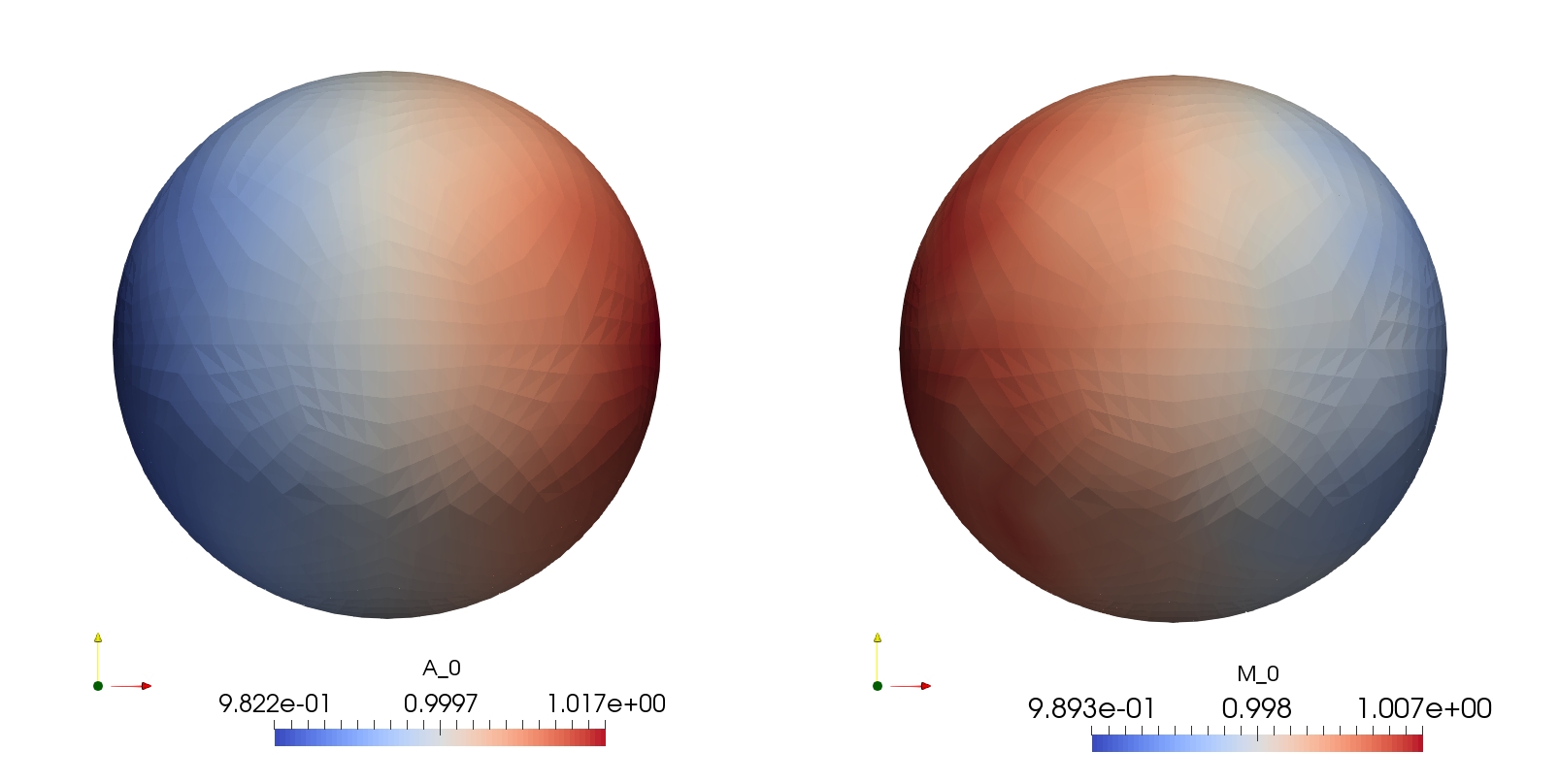}
 \caption{Graphical displays of the actin and myosin concentrations at time $t=1$. These are numerical solutions to the full system  \eqref{eq:fs} using the finite element formulation, as described in Section \ref{sec:selw11}. }\label{fig:fs_11}
\end{figure}

In this example, the actin and myosin concentration solutions will be the negation of each other with actin concentration highest on one side and myosin concentration highest on the opposite side. This mode is the first eigenfunction that one might hope to see for the organisation of actin and myosin in a cell because it is similar to what is often observed in a moving cell \cite{cooper2000,murrell2015}. $k^2_{1,1}=2.0816$ is also the lowest eigenvalue. Choosing parameters $\psi=20,\;c=-80,\;k_a=0.04,\;k_{ma}=0.05$ and $k_{am}=0.06$ and initial conditions 
\[
 a(\bx,0)=1+w_{1,1}^1(\bx)\times ran,\quad m(\bx,0)=1-w_{1,1}^1(\bx)\times ran,
\]
we observe that the mode $w_{1,1}^1$ is selected for actin and myosin. In Figure \ref{fig:fs_11} we plot the concentrations of actin and myosin at time $t=1$.  Blue indicates where the concentration is low, while red indicates that concentration is high. In this case very little deformation is seen.

\subsection{Cell deformation when \texorpdfstring{$w_{2,1}^0$}{w20} is excited initially}\label{sec:a20m20}
The simple first mode is not the only organisation which makes sense or shows similarities to organisation seen in cells. The cell can protrude in more than one direction because of actin accumulation at both ends, or deform in many other ways. Also, myosin could accumulate and ''squeeze'' on both sides. Therefore we continue by isolating other modes. We see that both the parameters, and the initial conditions, have an effect on which mode will grow. The first large deformation is seen when choosing initial conditions
\[
 a(\bx,0)=1+w_{2,1}^0(\bx)\times ran,\quad m(\bx,0)=1-w_{2,1}^0(\bx)\times ran.
\]
In Figure \ref{fig:a11m20} we plot the concentrations of actin and myosin and the displacement $\left(|\bU|=\sqrt{U^2+V^2+W^2}\right)$ at each point. The cell expands at the two ends where actin concentration is high and contracts in the middle where myosin concentration is high. So far the results are visually similar to the the results obtained by \cite{uduak} in the absence of myosin, one difference however is that there is only a very small volume increase because the cell is contracting in the middle as well as protruding. Other results (not shown) when the excited mode for myosin is the same  as the mode for actin are very similar to the previous model \citep{murphythesis2018}. Therefore, we investigate whether more interesting dynamics may occur if we try to excite differing modes for the two concentrations.

\subsection{Cell deformation when \texorpdfstring{$w_{1,1}^1$}{w11} and \texorpdfstring{$w_{2,1}^0$}{w20} are excited for actin and myosin, respectively}\label{sec:a11m20}
\begin{figure}[H]
\centering
\subfloat[t=0]{\includegraphics[trim = {0 0 0 40mm}, clip, width=10.5cm]{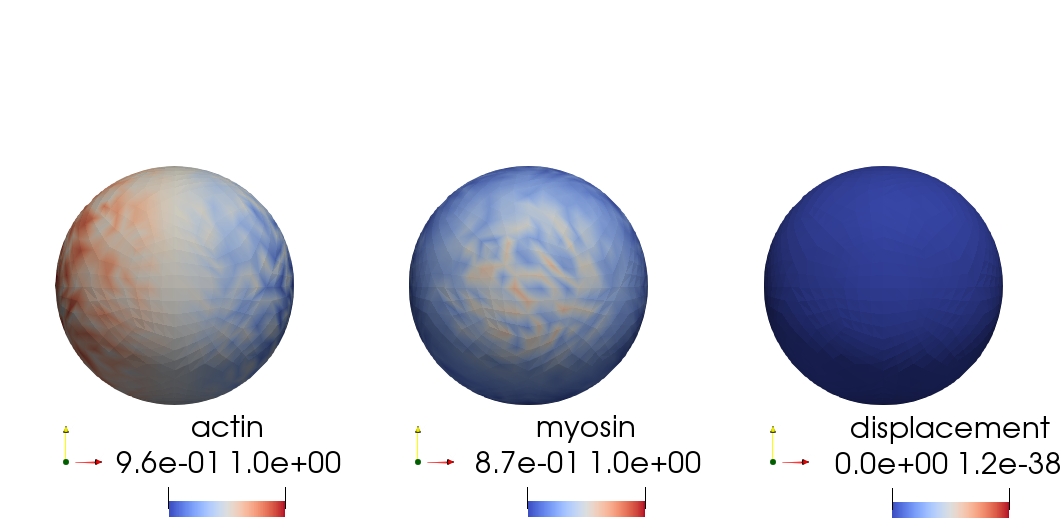}\label{fig:horns_t0}}\\
\subfloat[t=1]{\includegraphics[trim = {0 0 0 40mm}, clip, width=10.5cm]{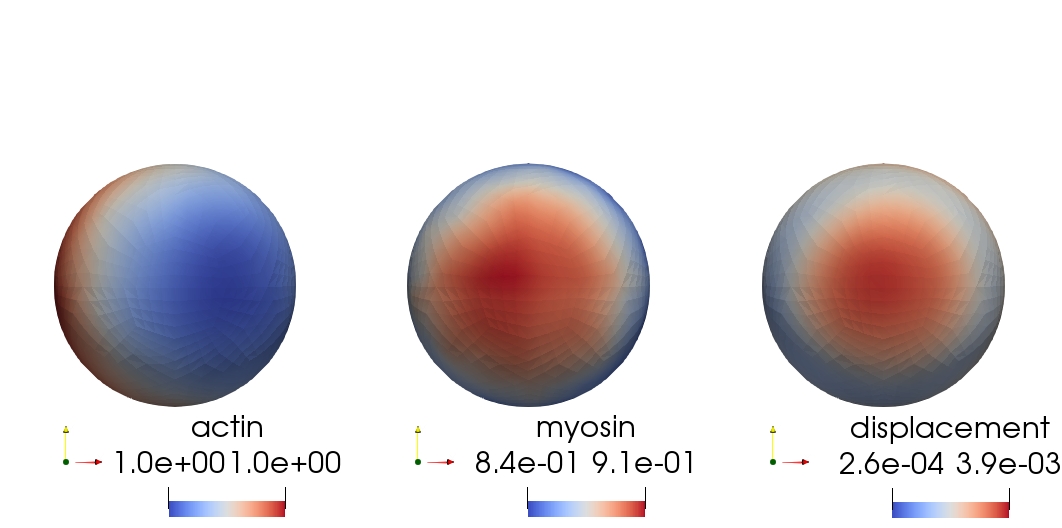}\label{fig:horns_t1}}\\
\subfloat[t=9]{\includegraphics[trim = {0 0 0 40mm}, clip, width=10.5cm]{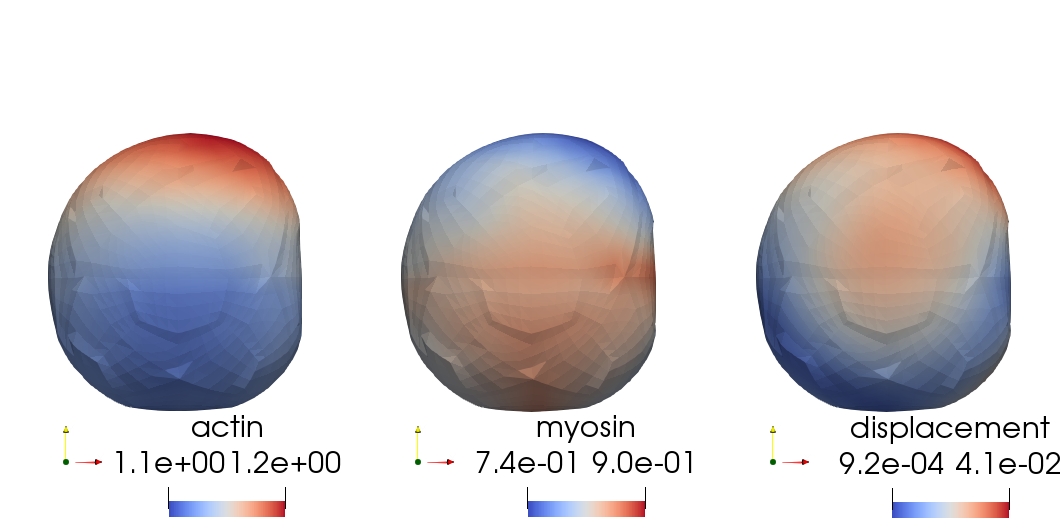}\label{fig:horns_t9}}\\
\subfloat[t=13]{\includegraphics[trim = {0 0 0 30mm}, clip, width=10.5cm]{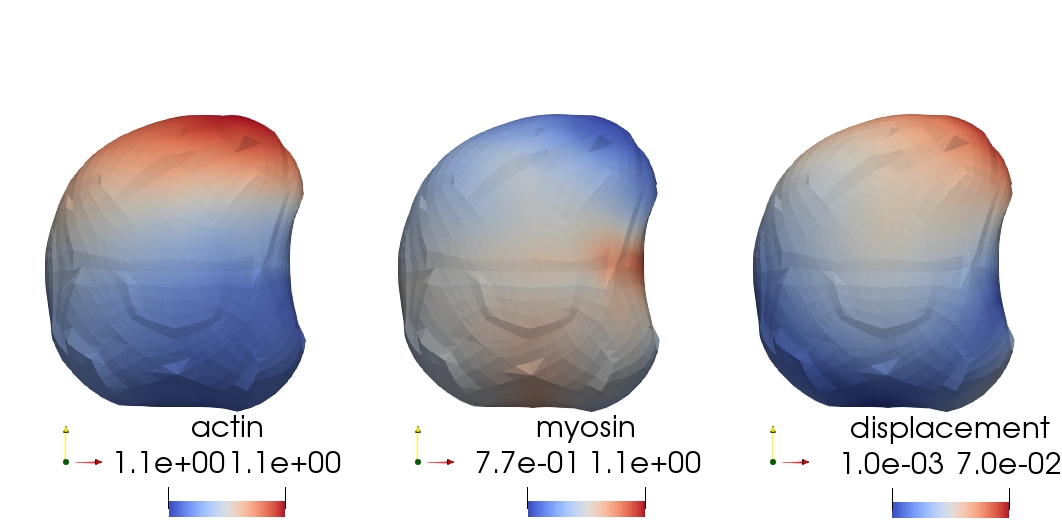}\label{fig:horns_t13}}
 \caption{Graphical displays of the actin and myosin concentrations, and the displacement at increasing time $t$, for the conditions described in Section \ref{sec:a11m20}. The sphere is squeezed where there is high myosin and then there is a protrusion in the area of high actin. Displacements in the $x,y$ and $z$ directions are shown in Figure \ref{fig:a11m20_minmax}. }\label{fig:a11m20}
\end{figure}

\begin{figure}[H]
 \centering
\includegraphics[width=10cm]{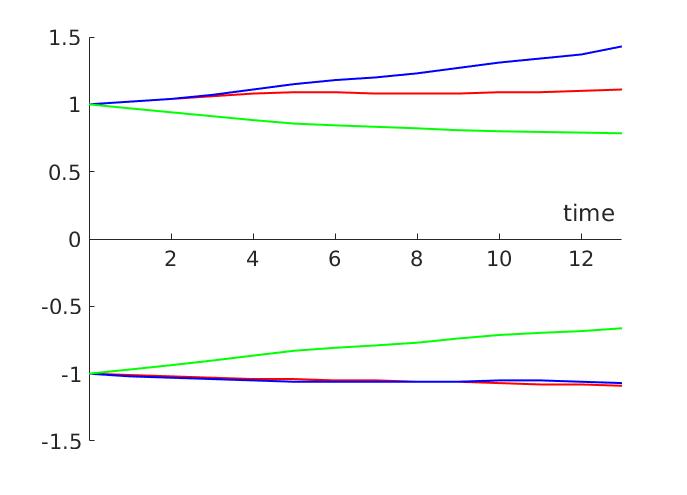}
\caption{Plot to show the minimum and maximum of $x$ (red), $y$ (blue) and $z$ (green) for the example in Section \ref{sec:a11m20} and Figure \ref{fig:a11m20}. The cell is contracting in the $y$ direction, expanding slightly in the $x$ direction but significantly in the positive $z$-direction. }\label{fig:a11m20_minmax}
\end{figure}

While the idea that actin and myosin accumulate in opposite sides is quite well founded, their concentration gradients are rarely exactly opposite. Therefore here we investigate if differing modes can be excited for actin and myosin. Choosing appropriate initial conditions, to encourage different modes to grow, we observe more irregular deformations. In Figure \ref{fig:a11m20} we plot the concentrations of actin and myosin and the displacement when the initial conditions are 
\[
 a(\bx,0)=1+w_{1,1}^1(\bx)\times ran,\quad m(\bx,0)=1+w_{2,1}^0(\bx)\times ran. 
\]
The cell squeezes where there is high myosin concentration and there is a protrusion in the direction of higher actin, this is also illustrated in Figure \ref{fig:a11m20_minmax} where the minimum and maximum in each spatial direction are plotted.

\subsection{Cell deformation when \texorpdfstring{$w_{2,1}^0$}{w20} and \texorpdfstring{$w_{1,1}^0$}{w11} are excited for actin and myosin, respectively}\label{sec:a20m11}
\begin{figure}[H]
\centering
\subfloat[t=0]{\includegraphics[trim = {7mm 5mm 3mm 30mm}, clip, width=8.1cm]{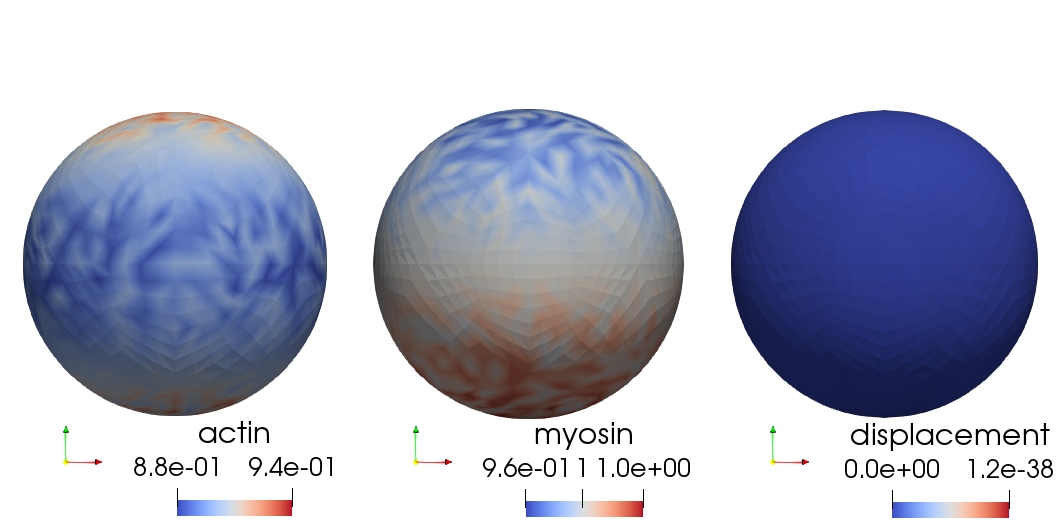}\label{fig:a20m11_t0}}\\
\subfloat[t=5]{\includegraphics[trim = {7mm 5mm 3mm 30mm}, clip, width=8.1cm]{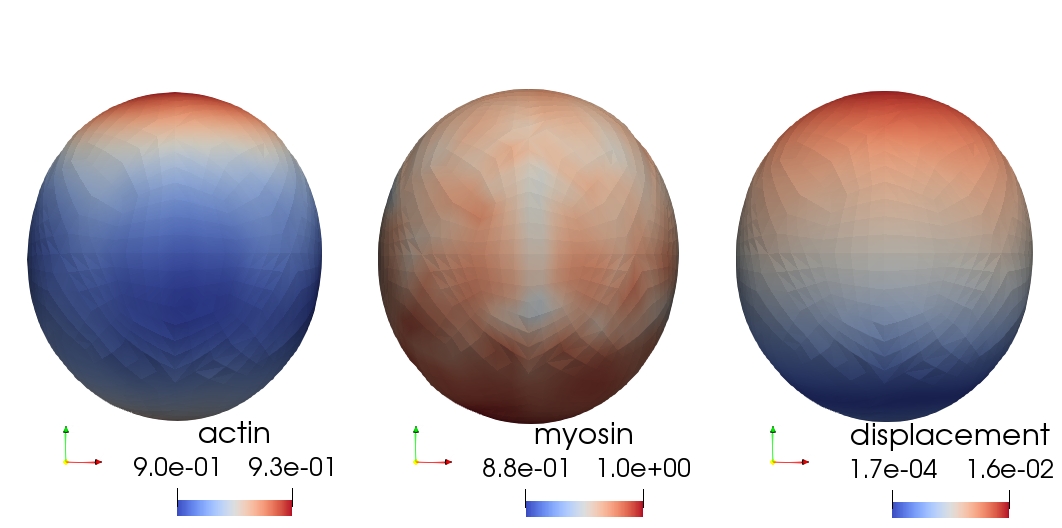}\label{fig:a20m11_t5}}\\
\subfloat[t=20]{\includegraphics[trim = {7mm 5mm 3mm 30mm}, clip, width=8.1cm]{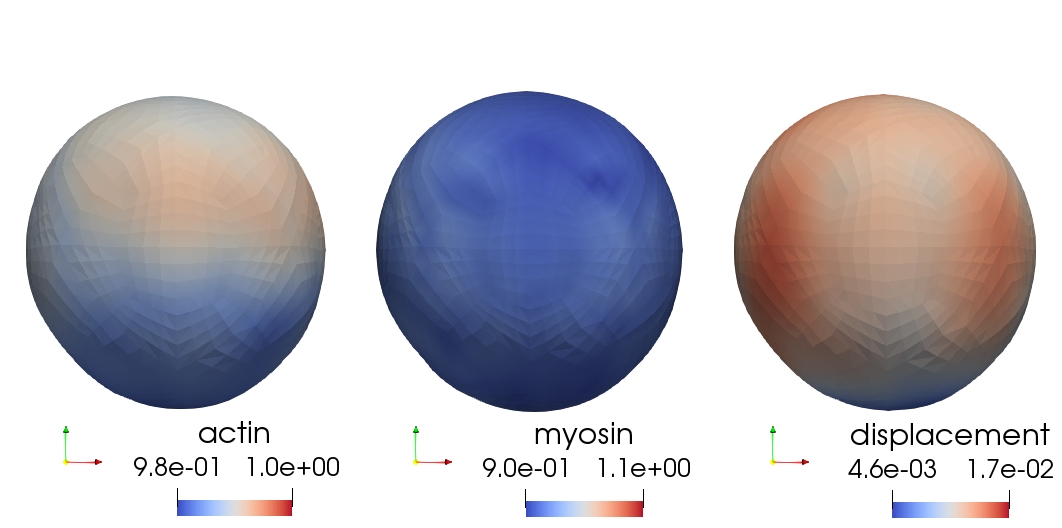}\label{fig:a20m11_t20}}\\
\subfloat[t=29]{\includegraphics[trim = {7mm 5mm 3mm 20mm}, clip, width=8.1cm]{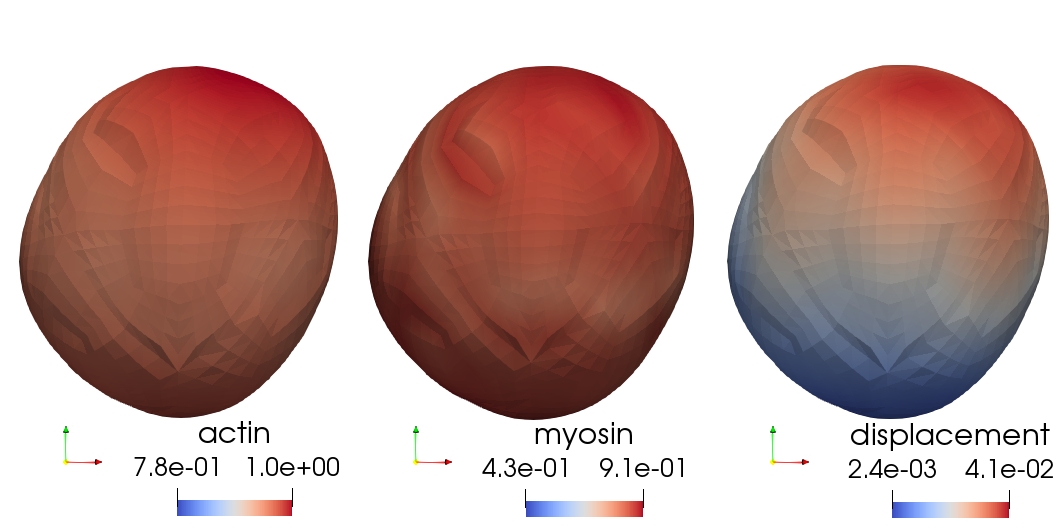}\label{fig:a20m11_t29}}\\
\subfloat[t=36]{\includegraphics[trim = {7mm 5mm 3mm 30mm}, clip, width=8.1cm]{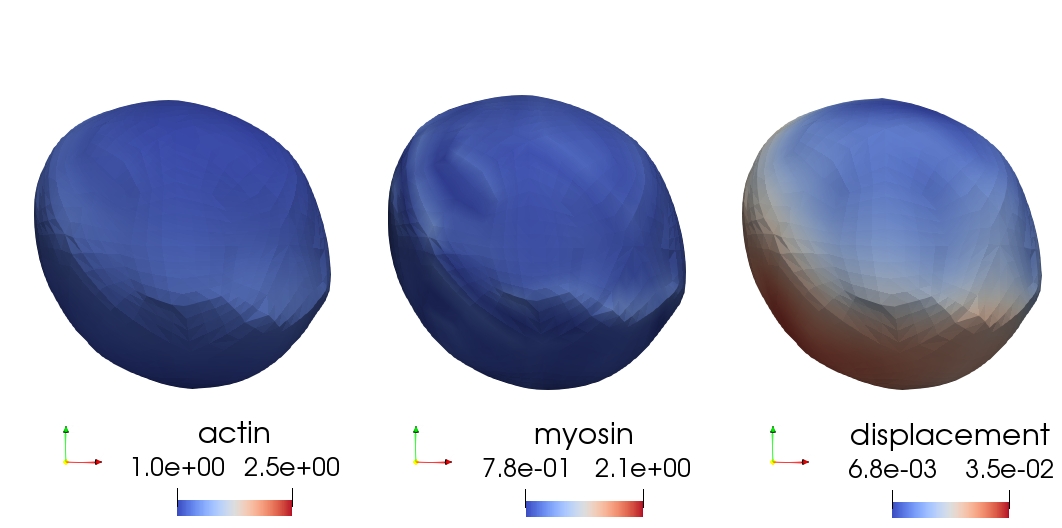}\label{fig:a20m11_t36}}
 \caption{Graphical descriptions of the solutions to simulations as described in Section \ref{sec:a20m11}. The cell expands and contracts twice, this can be seen more clearly in Figure \ref{fig:a20m11_lc}. The concentration of myosin inside the sphere is shown in Figure \ref{fig:a20m11_m}. }\label{fig:a20m11}
\end{figure}

\begin{figure}[H]
 \centering
\subfloat[t=5]{\includegraphics[trim = {0 0 0 30mm}, clip, width=3cm]{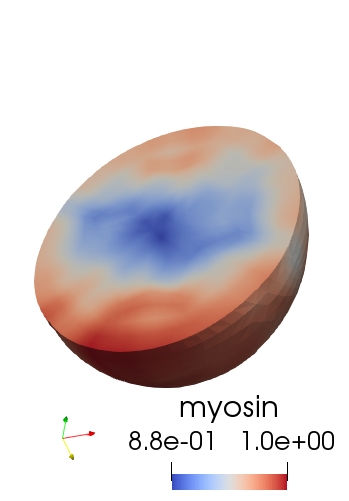}}\;
\subfloat[t=20]{\includegraphics[trim = {0 0 0 30mm}, clip, width=3cm]{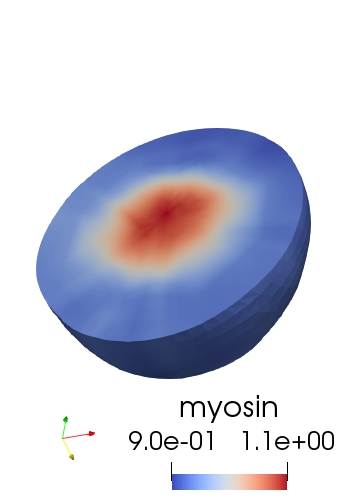}}\;
\subfloat[t=29]{\includegraphics[trim = {0 0 0 30mm}, clip, width=3cm]{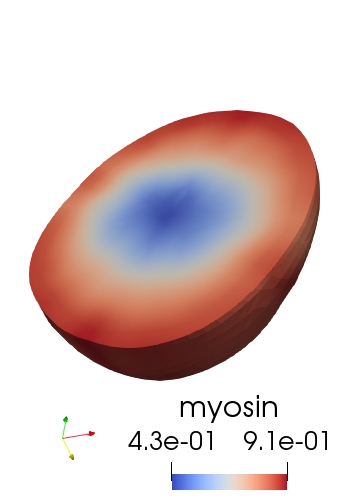}}\;
\subfloat[t=36]{\includegraphics[trim = {0 0 0 30mm}, clip, width=3cm]{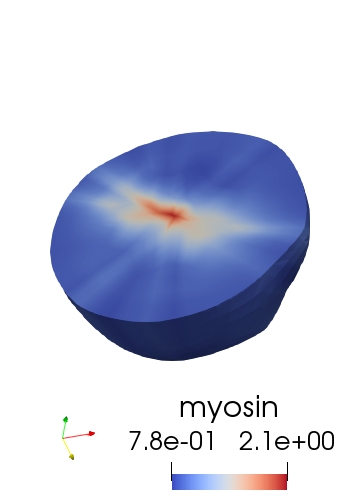}}
\caption{Graphical representations of solutions for myosin with a cut-through to see the behaviour in the bulk. When the cell is expanded the concentration is highest and the edge and later when it is contracted it is largest at the centre. This is also seen in a similar way for the actin concentration. }\label{fig:a20m11_m}
\end{figure}

\begin{figure}[H]
\centering
\subfloat[Length of cell in $z$-direction]{\includegraphics[width=7.3cm]{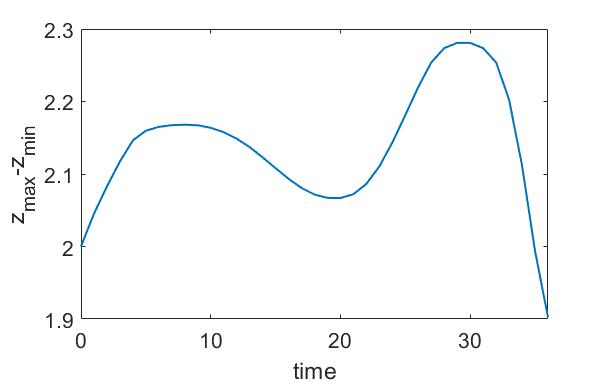}\label{fig:a20m11_zlength}}\;
\subfloat[Translation the centre of the cell in $z$-direction]{\includegraphics[width=7.3cm]{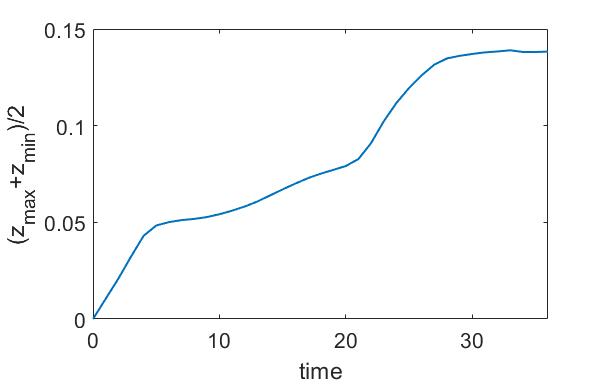}\label{fig:a20m11_cent}}
\caption{Plots to illustrate how the cell expands, contracts and translates in Figure \ref{fig:a11m20}. (Example \ref{sec:a20m11}). }\label{fig:a20m11_lc}
\end{figure}

In Figure \ref{fig:a20m11}, the initial conditions 
\[
 a(\bx,0)=1+w_{2,1}^0\times ran,\quad m(\bx,0)=1+w_{1,1}^0(\bx)\times ran,
\]
contain the same two eigenfunctions as the last example but with different orientations, ($w_{1,1}^0$ is a rotation of $w_{1,1}^1$), we observe a quite different deformation. There is high actin concentration at the top and bottom of the sphere. Without the effect of myosin one would expect it to extend in both directions in the same way as in Section \ref{sec:a20m20}, however there is high myosin at the bottom so the cell only protrudes upwards. Then at $t=5$ the protrusion slows and there is a contraction at the bottom where myosin concentration is high. There is another subsequent expansion and contraction with the actin and myosin concentrations reorganising to be is higher nearer the surface except when the cell is contracting, when the opposite is true, this is displayed in Figure \ref{fig:a20m11_m}. Figure \ref{fig:a20m11_cent} shows the translation of the cell and Figure \ref{fig:a20m11_zlength} shows the change in length in the $z$-direction. 

\subsection{Cell deformation when \texorpdfstring{$w_{1,1}^0$}{w10} and \texorpdfstring{$w_{3,1}^0$}{w30} are excited for actin and myosin, respectively}\label{sec:a11m30}
\begin{figure}[H]
\centering
\subfloat[t=0.1]{\includegraphics[trim = {0 5mm 0 40mm}, clip, width=10cm]{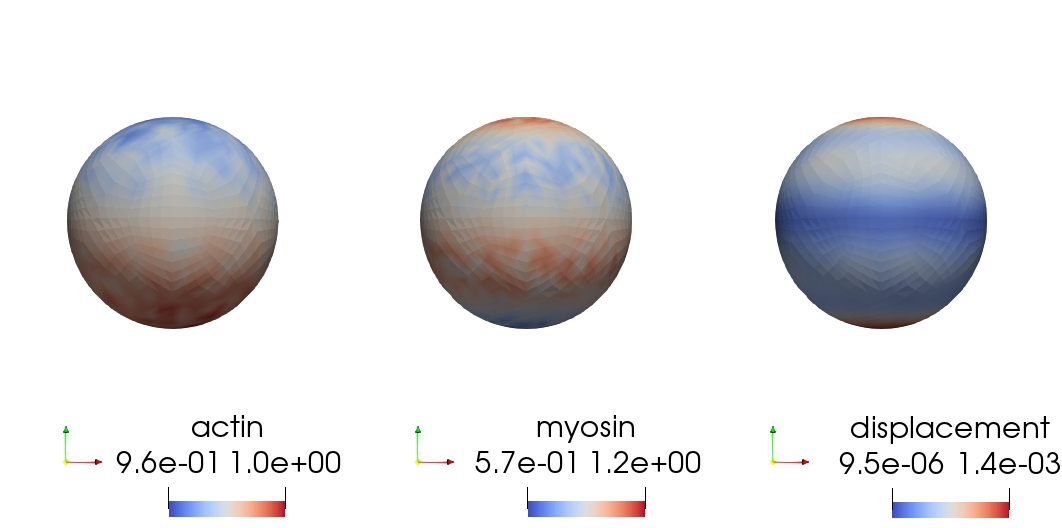}\label{fig:a11m30_t1}}\\
\subfloat[t=2]{\includegraphics[trim = {0 5mm 0 40mm}, clip, width=10cm]{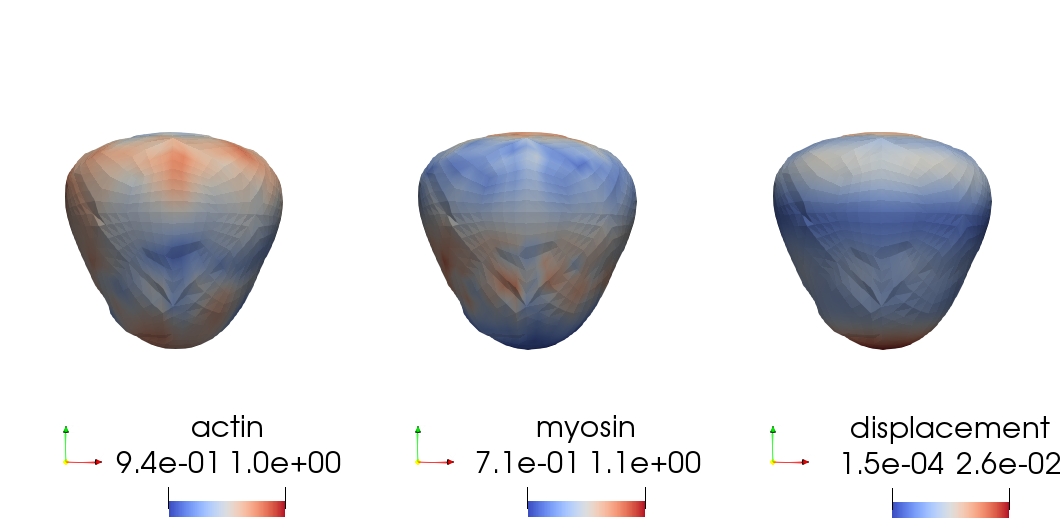}\label{fig:a11m30_t20}}\\
\subfloat[t=7.1]{\includegraphics[trim = {0 5mm 0 40mm}, clip, width=10cm]{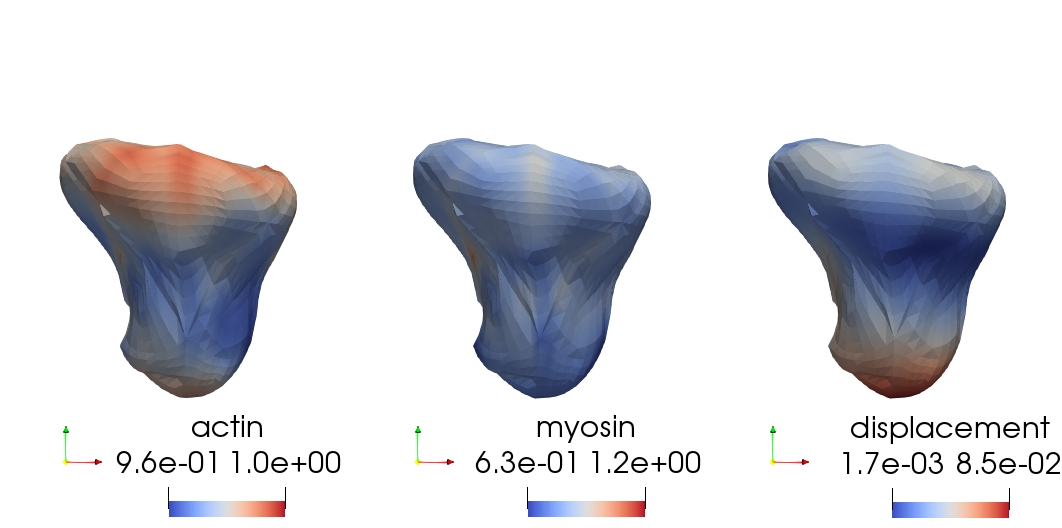}\label{fig:a11m30_t71}}
 \caption{Graphical displays of the solutions with conditions as described in Section \ref{sec:a11m30}. There is contraction in areas of high myosin, actin accumulates in areas of high curvature and the cell protrudes where there is high actin. }\label{fig:a11m30}
\end{figure}

\begin{figure}[H]
 \centering
\subfloat[Bounds on z]{\includegraphics[width=7cm]{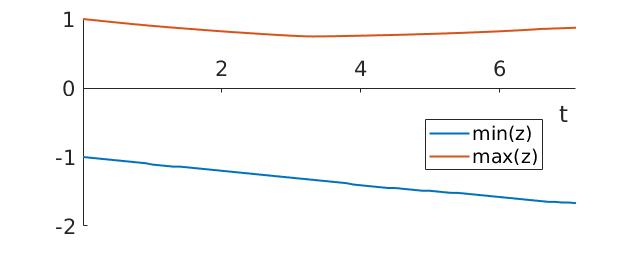}\label{fig:a11m30_z}}\;
\subfloat[Volume of the cell]{\includegraphics[width=7cm]{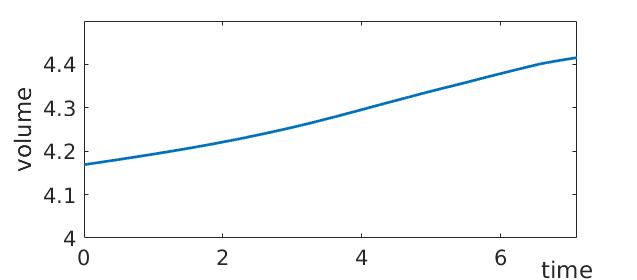}\label{fig:a11m30_vol}}
\caption{Plotting of the range demonstrates that there is a translation followed an expansion in the $z$-direction. The cell is also being squeezed in the $x$- and $y$-direction so we do not observe a significant volume increase. (Example \ref{sec:a11m30} and Figure \ref{fig:a11m30}). }\label{fig:a11m30_zvol}
\end{figure}

Next, we begin with initial conditions  
\[
 a(\bx,0)=1+w_{1,1}^1(\bx)\times ran,\quad m(\bx,0)=1+w_{3,1}^0(\bx)\times ran.
\]
This leads to a protrusion in the area with highest actin concentration which is pulling the cell in the negative $z$-direction. At the same time there is inward movement in areas of high myosin concentration. The cell has translated in the negative $z$-direction and this is plotted in Figure \ref{fig:a11m30}, and the change in volume is illustrated in Figure \ref{fig:a11m30_zvol}.

\subsection{Cell deformation when \texorpdfstring{$w_{1,1}^1$}{w11} and \texorpdfstring{$w_{4,1}^0$}{w40} are excited for actin and myosin, respectively}\label{sec:a11m40}
\begin{figure}[H]
\centering
\subfloat[t=0]{\includegraphics[trim = {0 5mm 0 40mm}, clip, width=10cm]{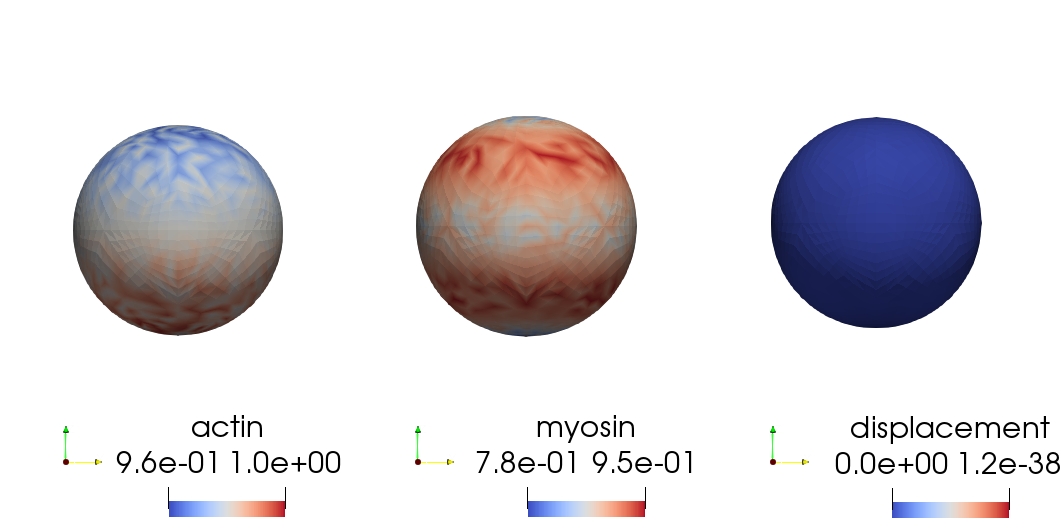}\label{fig:a11m40_t0}}\\
\subfloat[t=1]{\includegraphics[trim = {0 5mm 0 40mm}, clip, width=10cm]{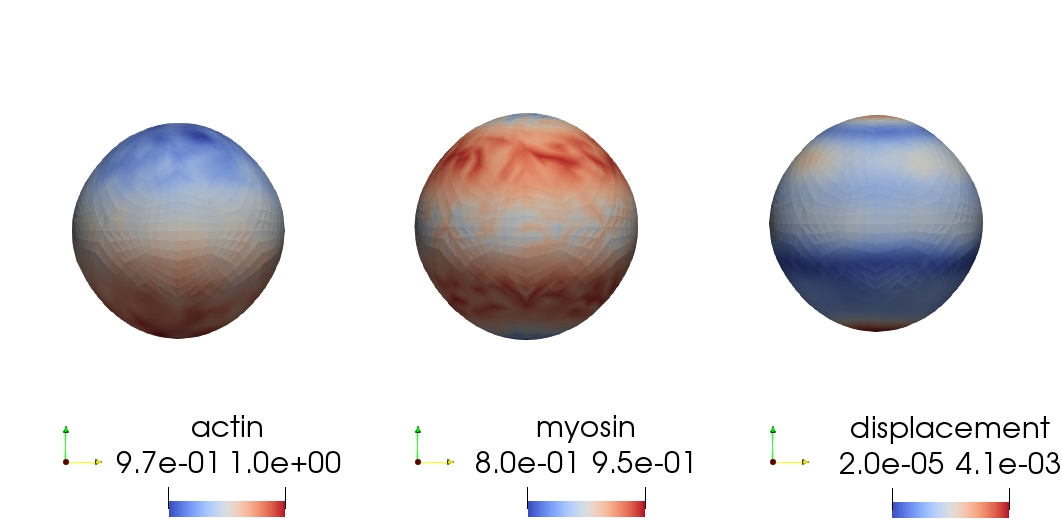}\label{fig:a11m40_t1}}\\
\subfloat[t=67]{\includegraphics[trim = {0 5mm 0 20mm}, clip, width=10cm]{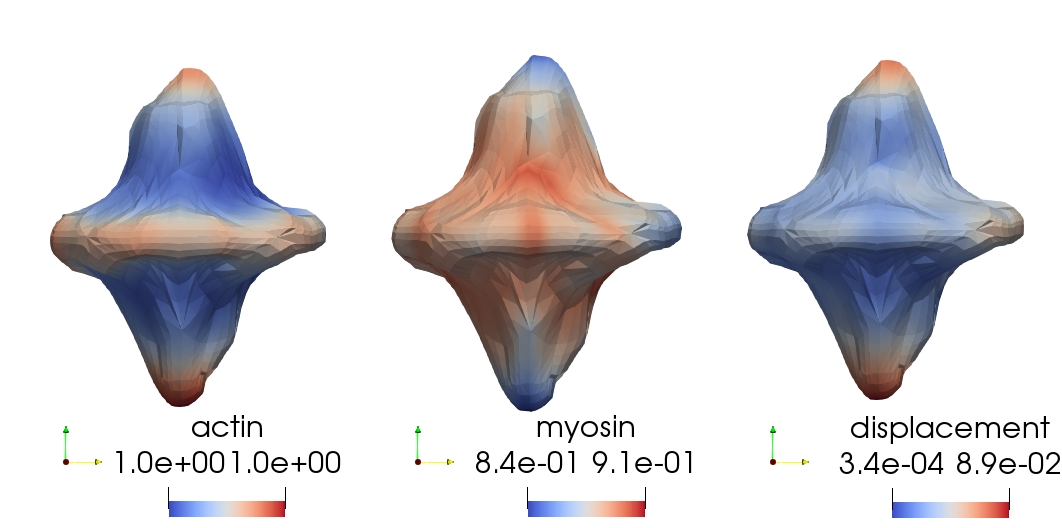}\label{fig:a11m40_t67}}
 \caption{Graphical displays of the solutions with conditions as described in Section \ref{sec:a11m40}. We see protrusions in a similar way to in Figure \ref{fig:a11m30} but in two directions. }\label{fig:a11m40}
\end{figure}

In another example of mixed modes, we start with 
\[
 a(\bx,0)=1+w_{1,1}^1(\bx)\times ran,\quad m(\bx,0)=1+w_{4,1}^0(\bx)\times ran.
\]
This leads to the expansion shown in Figure \ref{fig:a11m40}. The cell contracts inwards at areas of high myosin concentration and protrudes in the remaining areas, there are large protrusions in two opposing directions, the largest being the direction where actin was initially highest, subsequently, actin concentrates in areas of high curvature and protrudes further.
\begin{table}[H]
\centering
 \begin{tabular}{| l || c | | c | | c | | c | | c |}
  \hline
  Section & \ref{sec:a20m20} & \ref{sec:a11m20} & \ref{sec:a20m11} & \ref{sec:a11m30} & \ref{sec:a11m40} \\ \hline
  Figure & \ref{fig:a20m20} & \ref{fig:a11m20} & \ref{fig:a20m11} & \ref{fig:a11m30} & \ref{fig:a11m40} \\ \hline
  $\psi$ & 200 & 20 & 150 & 100 & 100  \\ \hline
  $c$ & -40 & -80 & -40 & -80 & -100  \\ \hline
  $k_{a}$ & 0.04 & 0.4 & 0.04 & 0.4 & 0.09   \\ \hline
  $k_{ma}$ & 0.05 & 0.5 & 0.05 & 0.05 & 0.09   \\ \hline
  $k_{am}$ & 0.06 & 0.12 & 0.06 & 0.07 & 0.15   \\ \hline
 \end{tabular}
\caption{Parameters for simulations in this section.}\label{table:fs_params}
\end{table}

\begin{figure}[H]
\centering
\subfloat[t=0]{\includegraphics[trim = {20mm 0 3mm 40mm}, clip, width=12cm]{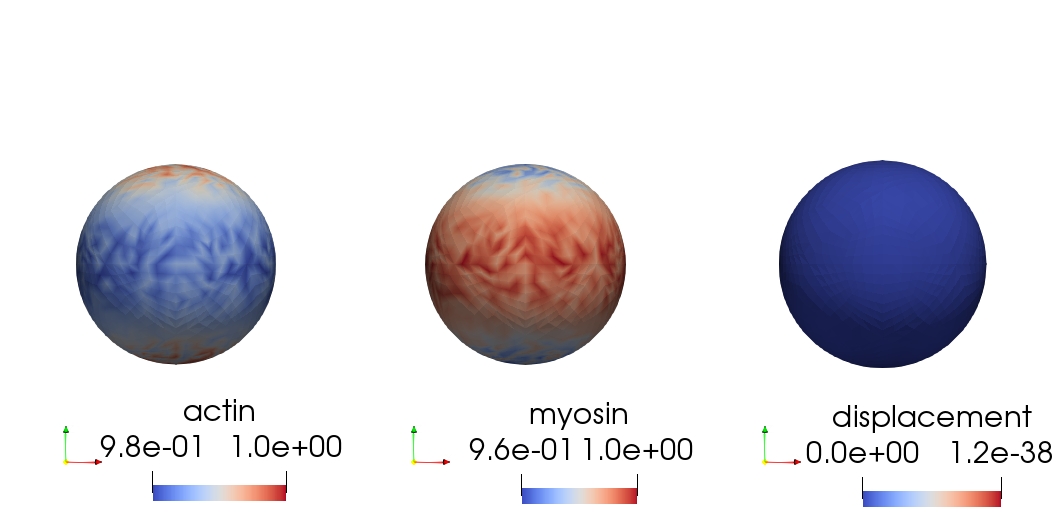}\label{fig:a20m20_t0}}\\
\subfloat[t=1]{\includegraphics[trim = {20mm 0 3mm 40mm}, clip, width=12cm]{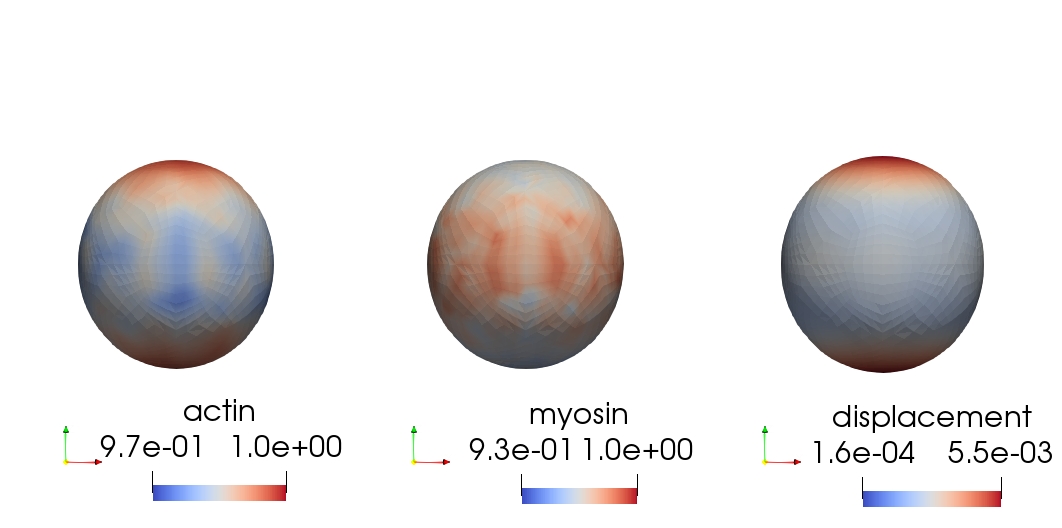}\label{fig:a20m20_t1}}\\
\subfloat[t=6.3]{\includegraphics[trim = {20mm 0 3mm 40mm}, clip, width=12cm]{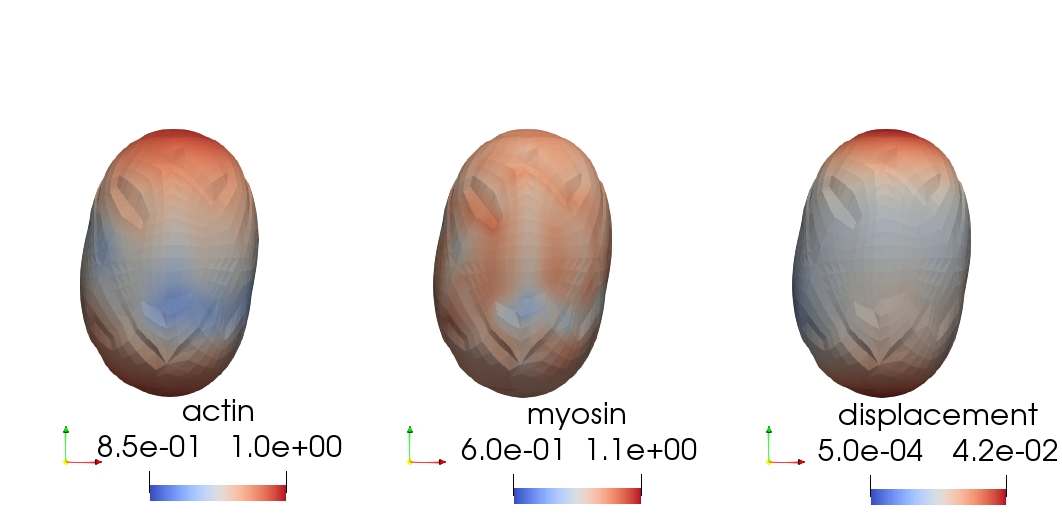}\label{fig:a20m20_t6-3}}
 \caption{Graphical displays of the actin and myosin concentrations, and the displacement at increasing time $t$, for the conditions described in Section \ref{sec:a20m20}. There is high actin at two ends, and high myosin in the middle. We then see in \protect\subref{fig:a20m20_t6-3} that the cell squeezes in the middle stretches in the two directions of higher actin. }\label{fig:a20m20}
\end{figure}




\subsection{\texorpdfstring{$L_2$}{L2} norms}
\begin{figure}
\centering
 \includegraphics[width=8cm]{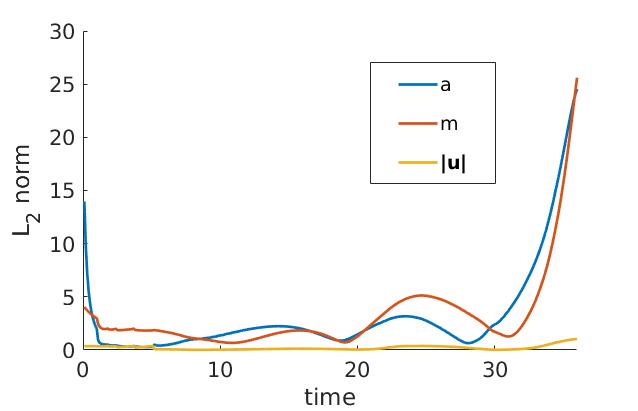}
\caption{Plot of the $L_2$ norm of difference between successive solutions for the example shown in Figure \ref{fig:a20m11}. There is an initial decrease due to diffusion, increases when the deformation is accelerating and decreases as deformation decelerates. The large error at the end is due to the mesh becoming very deformed and breaking. }\label{fig:l2_a20m11}
\end{figure}

In Figure \ref{fig:l2_a20m11} we plot the norm of differences between successive solutions in the case of the full system example in Section \ref{sec:a20m11}. We see an increase, or decrease, in the $L_2$ norm when the rate of deformation is accelerating, or decelerating, respectively. The qualitative changes in the $L_2$ norms are similar, but the changes in myosin and displacement appear slightly later than actin. This may suggest, in this example, that the change in actin triggers the change in the other variables.

In most of the numerical solutions in Section \ref{sec:sim_full} the cell becomes deformed in such a way that means the mesh becomes highly non-conforming and therefore the numerical method breaks down. In this case, re-meshing might be useful in order to allow the mesh to adapt to large deformations giving rise to a more stable numerical method. Equally important, adding a volume constraint term to the model might also prevent large expansions or contractions. These observations will form the subject of future studies.

\section{Conclusion}\label{sec:conclusion}
Our model revolves around an equation which balances elasticity, viscosity, contractility and pressure. Connected to this are two equations for the concentrations of F-actin and bound myosin. Unlike the previous study which our model is inspired by, (that of George \citep{uduak}), we used the software library deal.II \citep{dealii} to implement the finite element method. The key difference between this software and the previously used ALBERTA \citep{schmidtalberta} is that the elements are hexahedra rather than tetrahedra. The implementation is therefore different but we were able to appropriately replicate the previous results by George \cite{uduak} of cytomechanical model on a unit two dimensional disk (results not shown). Once this new implementation was verified we extended the model substantially in two ways: The two-dimensional formulation was extended to three dimensions. Unlike previous studies of this modelling framework, for the first time, we considered a second reaction-diffusion model to describe how myosin interacts with actin and how it contributes to cell contraction during cell migration. In the absence of experimental observations, we postulated hypothetical reaction kinetics describing the interaction between actin and myosin. As a first step in understanding solution behaviour in three dimensions of the full model, linear stability analysis close to bifurcation points was carried out and appropriate key parameter values were identified. An evolving finite element method was implemented in multi-dimensions.

Our numerical simulations showed that the model extends into three dimensions and the addition of myosin allows some symmetries to be broken and more striking deformations to emerge. In summary the main observations are:
\begin{itemize}
 \item There is outward movement in areas with high actin concentration, also where there is higher curvature, higher actin concentration is observed.
 \item This outward movement due to actin concentration is halted in areas with high myosin concentration. Additionally, if there is low actin and high myosin, we can see negative curvature.
 \item Identifying bifurcation parameters is more complicated than in previous models but effects of parameters can still be seen. The contractility due to myosin, $c$, strongly effects the speed of the deformation while the reaction constants $k_a,\;k_{ma}$ and $k_{am}$, and the diffusion coefficients $D_a$ and $D_m$ play a part in which mode the actin and myosin concentrations will arrange into. It is not as possible to isolate single modes just by picking parameter values, however, choosing initial conditions as a random number multiplied by the eigenfunction means it is possible to choose modes.
 \item Several examples show small translations.
 \item Initial conditions are a highly significant factor for the progression of the solutions. 
 \item In most examples the volume is increasing slightly but not significantly. Thus a mechanism for volume conservation or constraint could be a useful extension.
\end{itemize}

The modelling and computational framework presented in this article can readily be adapted to consider new experimentally driven reaction kinetics between actin and myosin or interactions between three or more molecular species, for example, studies using actin, myosin, GEF, Rho, Rac and CDC42 could be productive \citep{hall1998,holmes2016,nobes1995,simon2013}. Other useful extensions of the model could use or formulate re-meshing strategies.

\appendix
\section{Linear stability analysis}\label{sec:fsgc}
In the previous model, the reaction-diffusion equation alone could not cause patterning \cite{uduak}. Without the flow term, the prescribed reactions meant the concentration of actin would always return to the homogeneous steady state of $a=a_c$. In our case we have two coupled reaction-diffusion equations which are well known to induce patterning in certain cases.

We perform non-dimensionalisation to reduce parameters and simplify calculations. It also allows the reaction-diffusion equations to take the form necessary to use the standard conditions for diffusion driven instability. This is investigated in detail in \cite{murphythesis2018}. We choose the nondimensionalised parameters:
\begin{subequations}\label{eq:nondimparam}
\begin{align}
\tilde{t}=\frac{L^2}{D_a}t,\quad\quad \tilde{a}=\frac{a}{a_c}=a, \quad\quad \tilde{m}=\frac{m}{m_c}=m, \quad\quad d=\frac{D_m}{D_a},\quad\quad  \tilde{K}=\frac{K}{a_c},\quad\quad \gamma=\frac{L^2k_a}{D_a},  \\[10pt]
 \tilde{k_{ma}}=\frac{k_{ma}}{k_a}, \quad\quad \tilde{k_{am}}=\frac{k_{am}}{k_a}, \quad\quad \tilde{k_{m}}=\frac{k_{m}}{k_a}, \quad\quad \tilde{\Delta}=L^2\Delta,\quad\quad \tilde{\nabla}=L\nabla,\quad\; \tilde{\bu}=\frac{\bu}{L}, \quad\;\tilde{\phi}=\phi,\\[10pt] 
 \quad\;\tilde{\beps}=\beps,\quad\;\tilde{p}=p\frac{1+\nu}{E},\quad\;
\tilde{\bbeta}=\frac{\bbeta L}{D_a}, \quad\quad \tilde{a}_{sat}=\frac{a_{sat}}{a_c},  \quad\quad\tilde{\psi}=\psi a_c^2\frac{1+\nu}{E} \quad\quad \tilde{\mu}_i=\frac{\mu_iD_a(1+\nu)}{EL^2}.
         \end{align}
\end{subequations}
In the above, $L$ is the typical radius of a cell. Substituting appropriately and carrying out algebraic manipulations leads to the following non-dimensionalised system (for general kinetics)
\begin{subequations}
 \begin{align}
   \tilde{\nabla}\cdot\Bigg[(\tilde{\mu}_1\tilde{\beps}_t+\mu_2\tilde{\phi}_{t}\bI)+(\tilde{\beps}+\nu'\tilde{\phi}\bI)+ \sigma(\tilde{a})\bI+ \tilde{c}\tilde{m}\bI + \frac{\tilde{p}(\tilde{a})}{1+\tilde{\phi}}\bI\Bigg]&=0\\
   \pder{\tilde{a}}{\tilde{t}}+\tilde{\nabla}\cdot(\tilde{a}\tilde{\bbeta})-\tilde{\Delta}\tilde{a}-\gamma f(\tilde{a},\tilde{m})&=0\\
   \pder{\tilde{m}}{\tilde{t}}+\tilde{\nabla}\cdot(\tilde{m}\tilde{\bbeta})-d\tilde{\Delta}\tilde{m}-\gamma g(\tilde{a},\tilde{m})&=0.
 \end{align}
\end{subequations}
where $\sigma(\tilde{a})=\tilde{\psi}\tilde{a}^2 e^{-\tilde{a}/\tilde{a}_{sat}}$, $\nu'=\frac{\nu}{1-2\nu}$ and $\tilde{p}(\tilde{a})=\tilde{p}\big(1+\frac{2}{\pi}\delta(l)\arctan \tilde{a}\big)$. 
$f$ and $g$ have been nondimensionalised and we choose only functions such that system has a steady state at $(a_s,m_s,\bu_s)=(1,1,{\bf 0})$. Given small variations $\hat{a},\hat{m}$ and $\hat{\bu}$, consider the perturbation from the steady state $\tilde{a}=a_s+\hat{a}=1+\hat{a},\;\tilde{m}=m_s+\hat{m}=1+\hat{m},\;\tilde{\bu}=\bu_s+\hat{\bu}=\hat{\bu}$. This results in the linear system
\begin{subequations}
 \begin{align}
  \tilde{\nabla}\cdot\left[(\tilde{\mu}_1\hat{\beps}_t+\mu_2\hat{\phi}_{t}\bI)+(\hat{\beps}+\nu'\hat{\phi}\bI)+\hat{a}\sigma'(1)\bI+c\hat{m}\bI+\tilde{p}(1-\hat{\phi})\bI+\tilde{p}\frac{2}{\pi}\delta(l)\hat{a}\bI\right]=&{\bf 0},\\
  \pder{\hat{a}}{\tilde{t}}+\tilde{\nabla}\cdot(\hat{\beta})-\tilde{\Delta}\hat{a}-f_a\hat{a}-f_m\hat{m}=&0,\\
  \pder{\hat{m}}{\tilde{t}}+\tilde{\nabla}\cdot(\hat{\beta})-d\tilde{\Delta}\hat{m}-g_a\hat{a}-g_m\hat{m}=&0.
 \end{align}
\end{subequations}
We now look for solutions of the form 
\begin{equation}
 \hat{a}(\bx,t)=a^*e^{\lambda t+i\bk\cdot\bx},\quad\hat{m}(\bx,t)=m^*e^{\lambda t+i\bk\cdot\bx}\;\; \text{ and }\;\;\hat{\bu}(\bx,t)=\bu^*e^{\lambda t+i\bk\cdot\bx},
\end{equation}
 where $\lambda$ is the growth rate, $\bk$ is the wave vector, and $a^*,m^*$ and $\bu^*$ are small amplitudes. We require solutions to be non-trivial and so we obtain the stability matrix
\begin{subequations}
 \begin{align}
  \begin{vmatrix}
  \lambda+k^2-\gamma f_a & -\gamma f_m & \lambda i k \\ -\gamma g_a & \lambda+dk^2-\gamma g_m & \lambda i k \\ -ik\sigma'(1)-ik\tilde{p}\frac{2}{\pi}\delta(l) & -cik & \tilde{\mu}k^2\lambda+k^2(1+\nu')-\tilde{p}k^2
  \end{vmatrix}=0, \text{ where } k=|\bk| \\[10pt]
   \implies (h(\lambda):=)\mu\lambda^3 +a(k^2)\lambda^2+b(k^2)\lambda+c(k^2)=0,\label{eq:dispfull}\\[10pt]
  \text{where } a(k^2)=k^2(1+d)-\gamma(f_a+g_m)+1+\nu'-p-c-(\sigma'(1)+\tilde{p}\frac{2}{\pi}\delta(l)),\\
    b(k^2)=\tilde{\mu}(k^2-\gamma f_a)(dk^2-\gamma g_m)+(1+\nu'+p)(k^2(1+d)-\gamma(f_a+g_m))\\
     -c(k^2+\gamma(-f_a+g_a))+(\sigma'(1)+\tilde{p}\frac{2}{\pi}\delta(l))(\gamma(f_m+g_m)-dk^2)-\gamma^2\tilde{\mu}f_mg_a,\\
  \text{and } c(k^2)=(1+\nu'+p)\left((k^2-\gamma f_a)(dk^2-\gamma g_m)-\gamma^2f_mg_a\right).
 \end{align}
\end{subequations}
Thus $h(\lambda)=0$ \eqref{eq:dispfull} is our dispersal relation and we are concerned with the solution $\lambda$. There will be instability when $\Re(\lambda)>0$. We exploit this relation to isolate particular patterns/modes. The unstable modes will correspond to the eigenfunctions of the Laplacian on the sphere and $k^2$ the associated eigenvalues. 
 
\begin{figure}
\centering
 \includegraphics[trim = 100mm 2cm 50mm 60mm, clip,width=130mm]{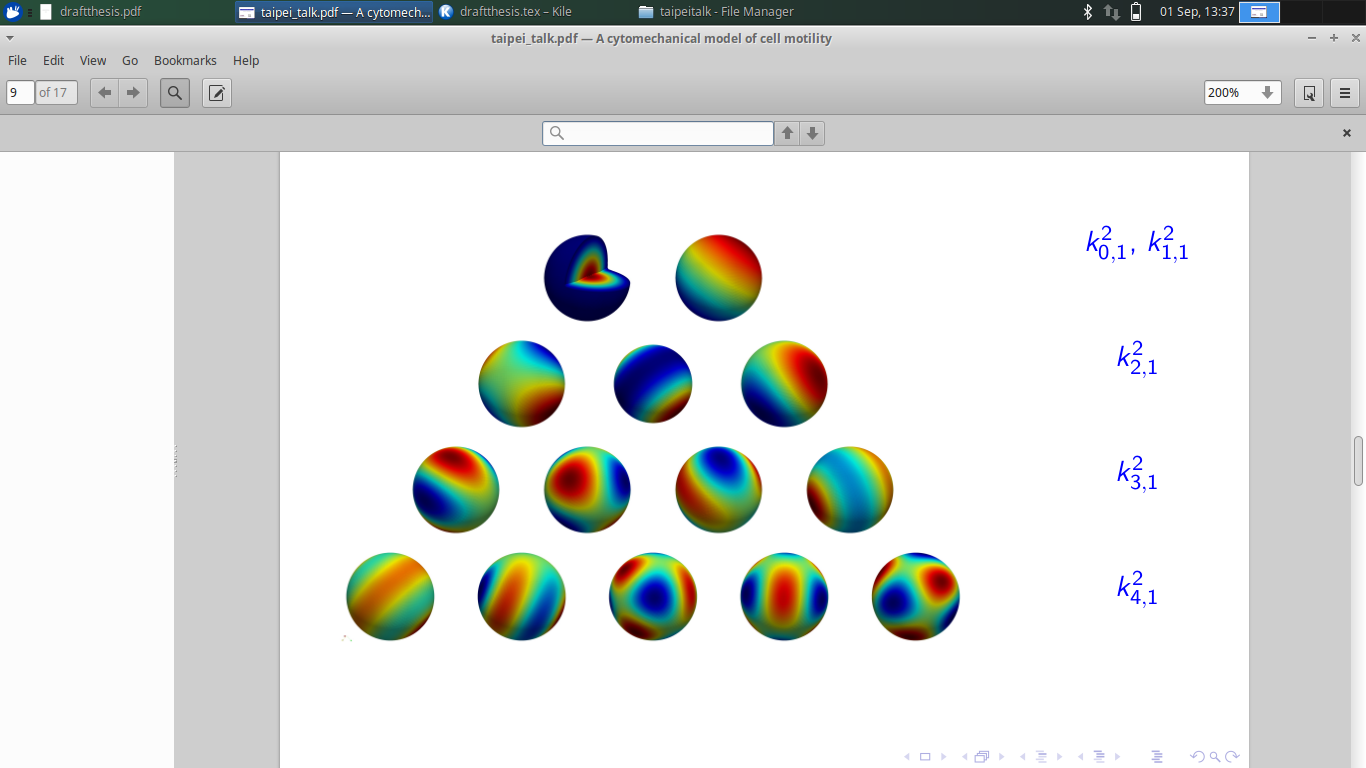}
 \caption{Analytical solutions to the eigenvalue problem on the unit sphere i.e. \eqref{eq:efsph} for selected values of $l,m,n$. For $l\geq 1$ there are multiple eigenfunctions for each eigenvalue. }\label{fig:sphereefs}
\end{figure}

\subsection{Eigenvalues and eigenfunctions of the Laplacian in the bulk of the unit sphere}
The eigenvalues on the unit sphere $\Omo=\{(x,y,z):x^2+y^2+z^2\leq 1\}$ (with homogeneous Neumann boundary condition) are well known and are obtained using separation of variables \citep{arfken,morimoto}. There are an infinite number of discrete solutions of the form
\begin{equation}\label{eq:efsph}
 \quad\quad\quad\quad w_{l,n}^m(r,\theta,\phi)=A_{l,n}^mJ_{l+\frac{1}{2}}(j'_{l+\frac{1}{2},n}r)e^{im\phi}P_l^m(\cos\theta),
\end{equation}
where $l,m,n$ are all integers such that $|m|\leq l \leq n,\;
A_{l,n}^m$ are constants, $J_\alpha(x)$ is a Bessel function of the first kind, i.e.
$J_\alpha(x)=\sum_{j=0}^\infty\frac{(-1)^j}{j!\Gamma(1+j+\alpha)}\left(\frac{x}{2}\right)^{2j+\alpha} $ with $ \Gamma(n)=(n-1)! $,
$ P_l^m(x)$ are associated Legendre polynomials and
  $j'_{l+\frac{1}{2},n}$ are zeros of the differential of the spherical Bessel function.
We can find the eigenvalues $k_{l,n}^2=(j'_{l+\frac{1}{2},n})^2$ numerically. It follows that for each eigenvalue $\lambda_{l,n}=k_{l,n}^2$ there are $2l+1$ possible eigenfunctions. Figure \ref{fig:sphereefs} shows the eigenfunctions for some selected values of $l$, $m$ and $n$. The wave numbers are discrete.

\subsection{Parameter selection}

The conditions on the positivity of the roots of \eqref{eq:dispfull} are numerous and the coefficients of the polynomial are burdensome. Therefore, we numerically find these roots and observe the real and imaginary parts.

\begin{figure}
 \centering
 \subfloat[Plot to show maximum real (solid line) and imaginary (dotted line) parts of the solution to the dispersal relation.]{\includegraphics[trim={0 0 0 2cm}, clip, width=80mm]{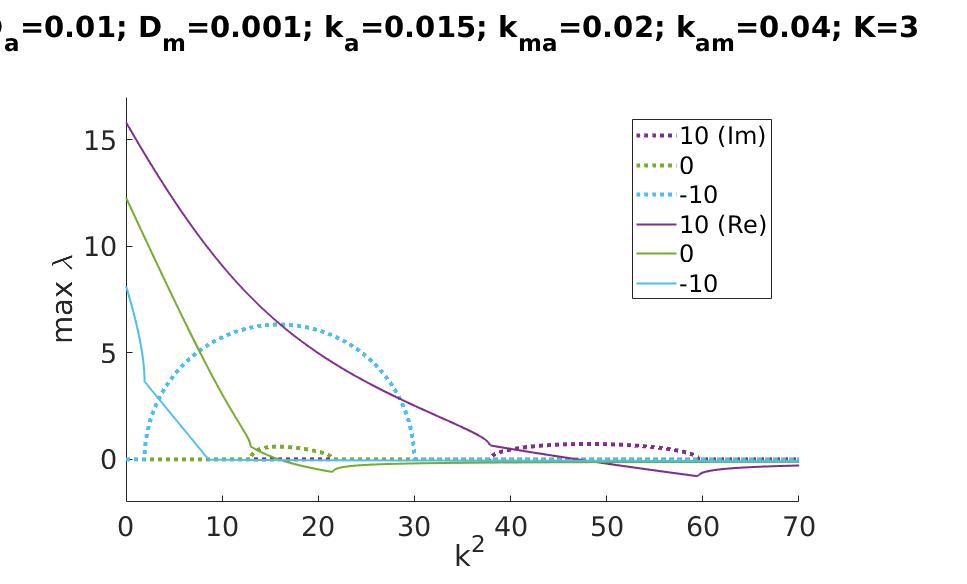}\label{fig:re_im_vary_c}}
\subfloat[Plot to show maximum real part of $\lambda$ as $\psi$ is varied.]
  {\includegraphics[trim={0 0 0 1cm}, clip, width=80mm]{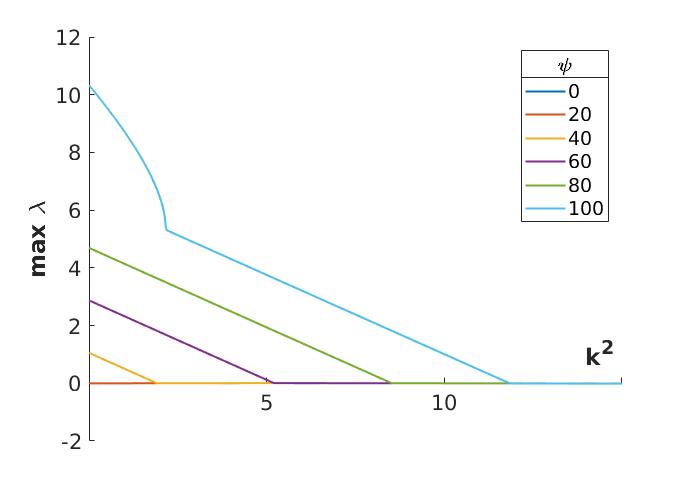}\label{fig:varypsi_full}}
  \caption{The effects of varying parameters on behaviour of solutions.}
\end{figure}

We found that contractility due to myosin ($c$), and due to actin ($\psi$) are particularly significant for finding unstable wavenumbers. In Figure \ref{fig:re_im_vary_c} we plot the real and imaginary parts of the solution against $k^2$ for three different values of c. We can see that when wavenumbers $k^2$ are less than $\sim 8.5$, then $\Re(\lambda)>0$ for the three values of $c$, therefore the wavenumbers will be unstable. Additionally these wavenumbers will be oscillatory for $c=10$. There are also regions of Hopf instability and oscillatory instability for all the three values of $c$. In Figure \ref{fig:varypsi_full} we fix other parameters and vary $\psi$ to see that, just like in \cite{uduak} higher values of $\psi$ mean higher wavenumbers can be excited.

\section*{Acknowledgements} 
This work (LM) was supported by an EPSRC Doctoral Training Centre Studentship through the University of Sussex. AM acknowledges support from the Leverhulme Trust Research Project Grant (RPG-2014-149). This work (AM) has received funding from the European Union Horizon 2020 research and innovation programme under the Marie Sklodowska-Curie grant agreement (No 642866). AM is a Royal Society Wolfson Research Merit Award Holder, generously supported by the Wolfson Foundation. LM acknowledges the support from the University of Sussex ITS for computational purposes. 

\end{document}